\documentclass[a4paper,11pt]{article}
\usepackage{graphicx}
\usepackage{dcolumn}
\usepackage{bm}
\usepackage{widetext}

\newcommand {\be} {\begin {equation}}
\newcommand {\ee} {\end {equation}}
\newcommand {\beq} {\begin {eqnarray}}
\newcommand {\eeq} {\end {eqnarray}}
\newcommand {\bc} {\begin {center}}
\newcommand {\ec} {\end {center}}

\def\disp {\displaystyle}

\def\Angs {A\raise11pt\hbox {\hskip-5pt\tiny$\circ$}}
\def\scs {\scriptsize}

\def\intl {\int\limits}
\def\ret {{\rm ret}}
\def\ph {{\rm ph}}
\def\ad {{\rm ad}}
\def\pl {{\rm pl}}
\def\nb {{\rm nb}}
\def\bb {{\rm bb}}

\def\rs {{\rm rs}}
\def\mt {{\rm mt}}

\newcommand {\sn}{\mathop {\rm sn}\nolimits}
\newcommand {\cs}{\mathop {\rm cs}\nolimits}
\newcommand {\sh}{\mathop {\rm sh}\nolimits}
\newcommand {\ch}{\mathop {\rm ch}\nolimits}

\def\d {{\rm d}}

\def\DR#1#2{\frac {\d #1}{\d #2}}

\def\cc {{\rm c}}
\def\g {{\rm g}}
\def\h {{\rm h}}
\def\o {{\rm o}}
\def\r {{\rm r}}
\def\t {{\rm t}}
\def\H {{\rm H}}
\def\O {{\rm O}}

\def\lim {{\rm lim}}
\def\aSB {a_{\rm SB}}
\def\Pl {{\rm Pl}}
\def\Hor {{\rm GHor}}
\def\Horr {{\rm KHor}}

\def\tl {\tilde {l}}

\def\e {{\hbox {\scs e}}}

\def\ff {\varphi}

\def\eps {\varepsilon}
\def\kB {k_{\rm B}}

\begin {document}

\bc
\title{\bf Basic geometric and kinematic features of the
Standard Cosmological Model}
\ec
\bc
\Large{\bf Basic geometric and kinematic features of the
Standard Cosmological Model}
\ec
\bc
{ D. I. Nagirner,$^1$, S. G. Jorstad$^{1,2}$, and A. V. Dementyev$^1$ }
\ec

\bc
$^1$ Astronomy Department, St. Petersburg State University,
Universitetskij Pr. 28, Petrodvorets, 198504 St. Petersburg, Russia

$^2$Boston University, 725 Commonwealth Ave., Boston, MA, 02215,
USA
\ec
\abstract{
We present a brief history of the construction of models of
the universe, followed by calculations of quantitative characteristics of basic geometric and
kinematic properties of the Standard Cosmological Model ($\Lambda$CDM). Using the Friedmann
equations of uniform space, we derive equations characterizing a $\Lambda$CDM model that describes a
universe corresponding to current observational data. The equations take into account the effects of
radiation and ultra-relativistic neutrinos. It is shown that the universe at very early and late
stages can be described to sufficient accuracy by simple formulas. Certain important moments of
cosmic evolution are determined: the times when densities of the gravitational components of the
universe become equal, when they contribute equally to the gravitational force, when the
accelerating expansion of space begins, and several others. The dependences of different distances
on redshift and the scale factor of space are derived. The distance to the sphere that expands with
the speed of light (the Hubble distance), and its current and future acceleration, are found.
Concepts of a horizon, second inflation, and second horizon are discussed. We consider the remote
future of the universe and the opportunity, in principle, of connection with extraterrestrial
civilizations.}

\section{Introduction} \label{sec:intro}

We present a brief history of the construction of cosmological models, starting with the models of
A. Einstein, V. de Sitter, A. Friedmann, and J. Lema\^{i}tre, and describe observational attempts to
choose model parameters such as the Hubble constant and the sign of curvature of space-time. We
follow the evolution of understanding of the role of the cosmological constant from its (1)
introduction by Einstein to compensate for gravitational attraction, (2) later near-elimination from
the theory, (3) association with some substance, and (4) recognition of this substance as the main
component of the universe, currently named ``dark energy,'' that defines the acceleration of the
expansion of space. All of these, along with ongoing technological progress, have led to the
formulation of the so-called ``Standard'' model, which is the model that most adequately describes
the existing universe.

We use the metric of Friedmann-Robertson-Walker and equations of Friedmann-Lema\^{i}tre, which form
the foundation of the majority of the models. We recall the definition of the ``critical density''
and discuss the densities of four noninteracting components of the universe: dark energy, dust
matter, which includes baryonic and dark matter, radiation, and ultrarelativistic neutrinos. We
derive the laws of evolution of the density of these components. Inclusion of these components allow
to obtain the general solution of the equations while highlighting the case of flat space-time.

We consider the problem of the propagation of radiation in the universe from a source to the
observer, a term of the geometrical horizon that arises in this case, and discuss different concepts
of distances in the universe. The relationship between a time derivative of the metric distance,
interpreted as the speed of expansion, and the distance itself --- the Hubble-Lema\^{i}tre law ---
and redshift is given. We explain why the cosmological redshift is not similar to the classic
Doppler effect.

We adopt the modern parameters of the Standard model: the temperature of the cosmic microwave
background (CMB), the temperature of neutrino gas associated with the CMB, the Hubble constant, and
a share of dark energy in the total density. These parameters allow us to calculate analytically the
evolution of the mass density of the four components and their relative contributions to the
critical density, which are employed to specify relations between scale characteristics of the model
and time. We derive approximate formulas of these relations at the initial and remote-future epochs
of expansion. The evolution of a role of the components at different epochs is analysed and values
of distances, speeds, and accelerations as functions of redshift are calculated. We estimate a
duration of the time interval needed to detect a change of the redshift and apparent luminosity of a
remote object with time.

The reasons for the existence of a second inflation and a second horizon are explained. Distances to
the horizons and to the place where the rate of expansion is equal to the speed of light (the Hubble
distance) are determined along with their speeds and accelerations. Finally, we discuss whether a
connection with extraterrestrial civilizations can be established in principle.

Based on the Standard model, the anisotropy of the cosmic microwave background, primary
nucleosynthesis, and the formation of large-scale structures in the universe are reproduced quite
accurately, as is presented in well known monographs, for example, \cite{ZeldNov, Narlikar, Thorn,
Weinberg, RubGorb}. Here we provide an overview of the geometric and kinematic properties of the
modern Standard cosmological model.

\section{Stages of building a model}

The path to the construction of a cosmological model, now known as the Standard model or
$\Lambda$CDM, was quite long. The history of this path is described in many cosmology textbooks.
Here we summarize the main stages of this history.

The intention to build a cosmological model, that is, a model of the entire universe, and not only
the solar system or the Galaxy, appeared as soon as in 1917 A. Einstein has formulated the equations
of the general theory of relativity \cite{Einshtcosm}, which describe gravitational fields
and the behavior of matter in them. He applied the equations to the homogeneous and isotropic
distribution of matter (following the ``cosmological principle'') and tried to find a stationary
solution of these equations to avoid problems of an origin and initial ultra-dense stages in the
history of the universe. To achieve this, he had to supplement the equations with the cosmological
term, so-called cosmological constant, or $\Lambda$ parameter \cite{Einshtcosm}, which corresponds
not to gravitational attraction, but to repulsion. The stationary solution was obtained only for a
closed universe and, moreover, as A. Eddington showed \cite{Eddington}, turned out to be unstable.

Another stationary solution was obtained by de Sitter \cite{deSitt} also for a closed
universe, but not containing matter (empty space). In this model, the passage of time at a certain
point of space depends on the distance to this position from the observer. The redshift of the
spectrum of a remote object was explained by the fact that a clock around it was running more slowly
than that of an observer on the Earth. According to this theory, the radius of the curvature of
space that does not change with time, $R=R_0$, is directly connected with the cosmological constant:
$\Lambda=3/R_0^2$. This model has prompted numerous works on the study of the "de Sitter World"
itself, in which properties of the symmetry of space in different coordinates (see the references in
\cite {Eddington}) were considered. In fact, these works are more of a mathematical character than
of cosmological insight, although they deal with some astronomical aspects and observed values
(speeds of galaxies, size of universe, etc.).

The first non-stationary solutions of the Einstein equations were obtained by a Russian
mathematician, fluid mechanician, and meteorologist A.A. Friedmann (1888--1925) for a matter without
pressure, uniformly distributed in space, first with a positive curvature \cite{Fridmanpos},
then with a negative curvature \cite{Fridmanneg}. In these works he also studied cases with
positive and negative values of the cosmological constant.

The Belgian theorist G. Lema\^{i}tre (1894--1966) also obtained non-stationary solutions
\cite{Lemaitre1} in 1927, not knowing about the works by Friedmann (when reprinting the article in
English, he mentioned the Friedmann work \cite{Fridmanpos}). Lema\^{i}tre was the first who included
radiation as a component of the universe and analytically derived the Hubble law ahead of its
observational discovery (now this law is called the Hubble---Lema\^{i}tre law). The idea of the Big
Bang at the beginning of the expansion of the universe belongs to Lema\^{i}tre as well
\cite{Lemaitre2}.

As for the cosmological constant, Einstein first doubted the need for its introduction
\cite{Einshtright}, and completely abandoned it when non-stationary solutions of the equations were
obtained \cite{Einshtref}. In 1929 E.~Hubble, using the 100-inch Mount Wilson Telescope, discovered
that the redshift of lines (translated by him into speed) in spectra of weak nebulae (which he had
found to be galaxies) increases with the distance to them. He interpreted this as recession. This
was considered as a proof of a non-stationarity of the universe. However, the values of the
parameters relevant to the universe remained uncertain and had to be determined by observations.

Observers turned to clarification of the cardinal question of whether the universe is closed or
open, that is, what is larger: the actual mass density of matter in space, $\rho$, or the critical
density, $\rho_\cc$, delimiting closed and open models. At that time it was believed that the
universe consists of primarily observable objects that radiate or absorb light, such as planets,
stars, galaxies, gas, and dust that form the baryonic component of the universe. Later, the dark
matter was added to the baryonic component, the amount of which is determined indirectly. This dark
matter is manifested through the attractive force of gravity: its existence explains the flat
rotation curves of spiral galaxies, the compactness of rich galaxy clusters, to which the virial
theorem is applicable, and gravitational lensing effect. The nature of this matter has not been
established, but its existence explains the observations.

Many observational works have been devoted to answering the cardinal question. These works use
cosmological tests that determine the ratio $\Omega_0 = \rho/\rho_\cc$ (or deceleration parameter
$q=\Omega_0/2$). The review by \cite{qdeterm} describes four different tests that allow one to
reconcile theory and observation by comparing the theoretical dependence of some quantity on
cosmological redshift $z$ with its observable behavior for different values of $\Omega_0$. One of
the tests suggests observations at different distances of a ``standard candle,'' that is, an object
(for example, a galaxy) of known intrinsic luminosity. For a long time, no such object was
available, since there is a variety of galaxy and quasar luminosities and the evolution of
luminosities is not fully understood.

The value of the Hubble constant $H_0$ is very significant, since it is used to determine the
critical density. The initial estimate of this constant by Hubble himself, $H_0=558$ km/s/Mpc, was
overestimated by almost ten times, because he mistook bright objects, identified later by radio
astronomy methods as HII regions, for stars, whereas the former are much brighter than the latter.
As a result, luminosities were overestimated, distances derived from them were underestimated, hence
the value of $H_0$ was overestimated, as established by A.~Sandage in 1958 \cite{Sanderror}. As time
went on, the Hubble constant was refined. This was the focus of a series of papers by
\cite{SandTamman}. The distance scale was based on the brightest blue and red giants in galaxies,
with reference to Cepheid variables. The results were summarized in a review \cite{H0determ},
although it was noted that different methods give values of $H_0$ that differ by several times due
to systematic errors of the methods. A detailed history of construction of cosmological models and
creation of the observational basis of cosmology is available in the book \cite{HoylBurbNar}.

The cosmological term was sometimes taken into account without clarifying its meaning; more often it
was ignored (see the review by \cite{Darkenerg}). The physical meaning to the cosmological term was
given by E.B. Gliner \cite{Gliner}. Considering various forms of non-traditional energy-momentum
tensors, he also interpreted the cosmological term as such a tensor corresponding to some substance,
characterizing the substance as vacuum-like. Currently this substance is called ``dark energy.'' Its
repulsive effect leads to models of exponential time-dependent expansion of the universe. In the
1980s, these solutions were used in the theory of cosmological inflation, which describes the
earliest stages of the evolution of the universe \cite{Guth,Linde}.

Substantial progress in determining cosmological parameters was achieved by technological advances
in the end of 20th century, which provided improved instruments for ground-based telescopes,
increased the sensitivity of receivers in different regions of the electromagnetic spectrum, and
made possible launches of measuring devices operating in space. As a result, observational
astrophysics, including cosmology, has become a multi-wavelength science. The decisive step towards
the most adequate model of the real universe was made by two groups that used as a standard candle
supernovae (SN) of type Ia, since the light curves of such SNs reliably determine their luminosity.
Group one \cite{Ries} using 10 SNs and group two \cite{Perl} using 42 SNs have found that the
main contribution to the critical density corresponds to the cosmological constant $\Lambda$. Both
papers mentioned above describe the history of employing SN of this type as a standard candle. For
example, this was done by \cite{SandTamman}, who used 16 SNs from the Coma Berenices and Virgo
clusters to construct a Hubble diagram, with the calibration based on two close supernovae, 1937c in
the galaxy IC4182 and 1954a in NGC4214. However, this was insufficient to make accurate conclusions.

Subsequently, the observations were extended to the highest redshifts, up to z$=1$ and soon to
z$=1.8$ \cite{Knop}, which confirmed the conclusion of \cite{Ries, Perl}, and rejected other
interpretations (see \cite{Darkenerg}). Confirmations were obtained by other types of observations
as well. Therefore, the conclusion by \cite{Ries, Perl} that $\Lambda$ plays a significant role in
the cosmos became a fundamental aspect of the Standard model that best describes the universe.
Although refinement of the parameters of the model continues, results of the Wilkinson Microwave
Anisotropy Probe (WMAP) \cite{WMAPparameters} and Plank \cite{Planck18112374,Planck18112376}
missions have allowed us to specify the value of $H_0$ with an accuracy of $10\%$ or even $7\%$.

\section{Basic equations of homogeneous cosmological models} \label{sec:basic}

In order to show the peculiarities of the Standard model we begin with a brief presentation of the
general theory of homogeneous cosmological models.

\subsection{Space-time metric} Any construction of a space-time model begins with the establishment
of its metric. For cosmological models the metric in conventional notation has the common form: $\d
s^2=c^2\d t^2-\d l^2$. It is convenient to determine the following alternative functions, related by
the equation $\cs_k^2\chi+k\sn_k^2\chi=1$: \be \label{eq:sncsdef} \sn_k\chi=\left\{ \begin
{array}{lll} \sin\chi & \hbox {{\rm for}} & k=1,  \\ \chi     & \hbox {{\rm for}} & k=0,  \\ \sh\chi
 & \hbox {{\rm for}} & k=-1, \\ \end {array}\right.\; \cs_k\chi=\left\{ \begin
{array}{lll} \cos\chi & \hbox {\rm for} & k=1,  \\ 1  & \hbox {\rm for} & k=0,  \\ \ch\chi  &
\hbox {\rm for} & k=-1. \\ \end {array}\right. \ee

Then the cosmological principle, according to which everything in the universe on large scale at
each moment is distributed uniformly and isotropically, is accepted, so that all points of space are
equal. Everything that occurs in each of them and with respect to them is completely the same. This
leads to the metric of space: \be \label{eq:dl2Retachi} \d
l^2=R^2(\eta)\left[\d\chi^2+\sn_k^2\chi\,\d\omega^2\right],\quad
\d\omega^2=\d\theta^2+\sin^2\theta\d\ff^2, \ee where $\eta$ and $\chi$ are dimensionless (conformal)
time and spatial coordinates, with $k$=1,0,$-1$ corresponding to closed, flat, and open space,
respectively, $R(\eta)$ is the radius of curvature, and $\d\omega^2$ is the metric on a sphere of
unit radius, with angular coordinates $\theta$ and $\ff$. The curvature, $k/R^2$, is constant at
every moment over entire space. The only point of space that needs to be specified is the location
of the observer (humanity), and the reference time, which is the current epoch.

Let us fix a moment $t$ and the corresponding conformal time $\eta$ and draw a ray with the
beginning at the position of the observer O, with direction determined by spherical angles
$\theta=\theta_0,\,\,\ff=\ff_0$. There is a point on the sphere with a radius of unity which
corresponds to this ray such that the distance $\d\omega=0$. Let us choose a point $P$ on the ray at
distance $l$ from point O and with a spatial coordinate $\chi$, and draw a sphere, all points of
which are at a distance $l$ from point O, so that it passes through the selected point $P$. Note
that in the general case (if $k\ne 0$) the point O is not the center of the sphere (let us call it a
quasi-center), the radius of the sphere, $r$, is not equal to $l$, and the ray is not a straight
line. (In analogy, the radius of a parallel on the Earth’s surface is not equal to the distance
between the parallel and pole; the latter is not the center of the parallel, and a ray goes along
the meridian, which is not a straight line.) The connection of $l$ and $r$ with a radius of
curvature is expressed by the following equations: \be \label{eq:lmetrRetar} l=R(\eta)\chi,\quad
r=R(\eta)\sn_k\chi, \ee wherein $l>r$ at $k=1$, $l<r$ at $k=-1$,  and $l=r$ at $k=0$.

The adoption of a space metric defines all geometric properties of space. For example, the volume of
space in a sphere whose points lay on the distance $l=R(\eta)\chi$ from O ($\chi=0$). A radius of the sphere 
is $R(\eta)\sn_k(\chi_0)$ for $k\ne 0$ is equal to 
\be
V(\eta,\chi)=R^3(\eta)4\pi\intl_0^{\chi_0}\sn_k^2\chi\d\chi= R^3(\eta)2\pi\left|\chi_0-\frac
{1}{2}\sn_k(2\chi_0)\right|. \ee 
For $k=1$ the total volume of the space, $2\pi^2R^3(\eta)$, is
finite, since $\chi\leq\pi$. For $ k=0 $, it is formally necessary to replace $\sn_k(2\chi_0)$ by
its Taylor series representation up to the $\chi_0^3$ term.

The conformal time $\eta$ is connected with the usual time, which is fixed by the value of the
density, by the equality $c\d t=R(\eta)\d\eta$. In these variables, the metric takes the form of the
Friedmann-Robertson-Walker metric: \be \label{eq:FRWmetr} \d
s^2=R^2(\eta)\left[\d\eta^2-\d\chi^2-\sn_k^2\chi\, \d\omega^2\right]. \ee

\subsection{The Friedmann Equations}

From the equations of the Einstein theory of gravitation (GR) with the metric (\ref{eq:lmetrRetar}),
two equations for the radius of curvature can be derived, known as the Friedmann equations: \beq
\label{eq:ddotR} & \strut\disp \ddot {R}=-\frac {4\pi G}{3}\left(\rho+3\frac {P}{c^2}\right)R+ \frac
{\Lambda c^2}{3}R, & \\ \label{eq:dotR2} & \strut\disp \dot {R}^2=\frac {8\pi G}{3}\rho R^2+\frac
{\Lambda c^2}{3}R^2- kc^2. & \eeq Here $\rho$ is the total mass density of matter and radiation, $P$
is their total pressure, and $\Lambda$ is the cosmological constant.

The terms with the cosmological constant in Eqs. (\ref{eq:ddotR})--(\ref{eq:dotR2}) can be appended
to the first term in each expression, which allows us to determine the total mass density and
pressure, as well as the gravitational mass density: \be \label{eq:rhoPtdef}
\rho_\t=\rho+\rho_\Lambda,\quad P_\t=P+P_\Lambda,\quad \rho_\g=\rho_\t+3\frac {P_\t}{c^2}. \ee To
satisfy these relations, it is necessary to determine the density and pressure corresponding to the
cosmological term as follows: \be \label{eq:rhoPLambdadef} \rho_\Lambda=\frac {\Lambda c^2}{8\pi
G},\quad P_\Lambda=-\frac {\Lambda c^4}{8\pi G}. \ee It is negative pressure that produces
repulsion.

Then the equations can be written in shorter form: \beq \label{eq:ddotRtot} & \strut\disp \ddot
{R}=-\frac {4\pi G}{3}\rho_\g R, & \\ \label{eq:dotRtot} & \strut\disp \dot {R}^2=\frac {8\pi
G}{3}\rho_\t R^2-kc^2, & \eeq or, for a scale factor $\disp a=\frac {R}{R_0}=\frac {1}{1+z}$, \be
\label{eq:eqsscalfact} \ddot {a}=-\frac {4\pi G}{3}\rho_\g a,\,\, \dot {a}^2=\frac {8\pi
G}{3}\rho_\t a^2-\frac {kc^2}{R_0^2}. \ee The past corresponds to the values $a<1$; at the current
epoch $t=t_0$, $R=R_0,\,a=1$, and redshift $z=0$; and for the future $a>1$, $-1<z<0$. According to
its definition, the scale factor is tied to the current epoch.

For compatibility of the equations, an additional condition is required: \be \label{eq:condmut} \dot
{\rho}_\t=-3\left(\rho_\t+\frac {P_\t}{c^2}\right)H, \ee where the new variable, $\disp
H=\frac{\dot{R}}{R}$, like the radius of curvature, depends only on time. We call it the ``Hubble
parameter.'' Its current value, $H_0$, is the Hubble constant. The relation (\ref{eq:condmut}) can
be interpreted as the condition of adiabatic expansion of space along with its contents, since it
implies that the differential of the total energy in volume $V$ satisfies \be \label{eq:adiabat}
\d(c^2\rho_\t V)=-P_\t\d V. \ee Equation (\ref{eq:ddotRtot}) also yields an equation that the Hubble
parameter obeys: \be \label{eq:dotHrho} \dot {H}=-H^2-\frac {4\pi G}{3}\rho_\g. \ee

\subsection{Non-interacting components}

After the annihilation of electron-positron pairs, the composition of the universe became simpler,
and since then its components have been the non-relativistic matter (including baryons and dark
matter), radiation, neutrinos, and so-called dark energy (formerly called the vacuum), whose density
and pressure are given by formulas (\ref{eq:rhoPLambdadef}). Of course, at high temperatures the
matter was relativistic, but then its abundance was small. Similarly, due to the finiteness of the
mass, at low temperatures neutrinos transformed from ultrarelativistic to relativistic (moderately
or weakly) or even non-relativistic, but by then their mass fraction was small and difference of
their masses from zero do not affect evolution of the universe \cite{DINTur}. Therefore, we assume
that during the entire evolution of the universe over the period under consideration, the matter has
not exerted pressure. This means that the matter has been non-relativistic (dust-like), while all
kinds of neutrinos can be treated as ultra-relativistic.

We can assume that during this period the four components did not interact with each other. Dark
energy in general does not interact with anything, while the interaction of cosmological neutrinos
with matter essentially ceased before the annihilation epoch. Radiation interacted with matter,
namely, free electrons and photons interacted until the end of the recombination epoch. However,
after annihilation and establishment of equilibrium distributions, Compton (Thompson) scattering
changes significantly neither the number of photons and electrons nor their energies. Therefore, the
evolution of the components took place independently thereafter.

In view of the foregoing, the equations of state of the four indicated components: the dust matter
(d), the radiation (r), neutrinos ($\nu$), and dark energy ($\Lambda$) are written in the form \be
\label{eq:eqsstate} P_\d=0,\quad P_\r=\frac {c^2}{3}\rho_\r,\quad P_\nu=\frac {c^2}{3}\rho_\nu,
\quad P_\Lambda=-c^2\rho_\Lambda. \ee The condition (\ref{eq:condmut}) is fulfilled for each
non-interacting component separately: \be \label{eq:separatecond} \dot {\rho}_\d=-3\rho_\d
H,\quad\dot {\rho}_\r=-4\rho_\r H,\quad \dot {\rho}_\nu=-4\rho_\nu H,\quad\dot {\rho}_\Lambda=0. \ee
The equations are easily integrated, which provides the evolution of the densities of the
components: \be \label{eq:evoldens} \rho_\d=\frac {\rho_\d^0R_0^3}{R^3}=\frac
{\rho_\d^0}{a^3},\; \rho_\r= \frac {\rho_\r^0R_0^4}{R^4}=\frac {\rho_\r^0}{a^4},\;
\rho_\nu=\frac {\rho_\nu^0R_0^4}{R^4}=\frac {\rho_\nu^0}{a^4},\; \rho_\Lambda=\rho_\Lambda^0. \ee
Here, as above, the index 0 means belonging to the current epoch.

\subsection{Critical parameters}

In theory, the critical density, which plays an important role, and the fraction of all components
in it are defined as: \be \label{eq:critdens} \disp\rho_\cc=\frac {3H^2}{8\pi G},\quad
\rho_\t-\rho_\cc=k\frac {3c^2}{8\pi GR^2},\quad \Omega_\t=\frac
{\rho_\t}{\rho_\cc},\quad\Omega_\t-1=k\frac {c^2}{\dot {R}^2}. \ee The sign of differences
$\rho_\t-\rho_\cc$ and $\Omega_\t-1$ coincides with the sign of $k$. If $\rho_\t-\rho_\cc=0$ then
$\Omega_\t-1=0$ and $k=0$. The shares of individual components are also determined: \be
\label{eq:OmegadrnuLtdef} \Omega_\d=\frac {\rho_\d}{\rho_\cc},\; \Omega_\r=\frac
{\rho_\r}{\rho_\cc},\; \Omega_\nu=\frac {\rho_\nu}{\rho_\cc},\; \Omega_\Lambda=\frac
{\rho_\Lambda}{\rho_\cc},\; \Omega_\t=\Omega_\d+\Omega_\r+\Omega_\Lambda. \ee The densities of
the components are expressed in terms of the current critical density and their current shares in
it: \be \label{eq:rhoaOmeg} \rho_\d=\rho_\cc^0\frac {\Omega_\d^0}{a^3},\,\, \rho_\r=\rho_\cc^0\frac
{\Omega_\r^0}{a^4},\,\, \rho_\nu=\rho_\cc^0\frac {\Omega_\nu^0}{a^4},\,\,
\rho_\Lambda=\rho_\cc^0\Omega_\Lambda^0,\,\, \rho_\cc^0=\frac {3H_0^2}{8\pi G}. \ee Since radiation
and neutrinos evolve in the same way, one can introduce their common density and pressure: \be
\label{eq:rhornudef} \rho_{\r\nu}=\rho_\r+\rho_\nu=\rho_\cc^0\frac {\Omega_{\r\nu}^0}{a^4},\,\,
P_{\r\nu}=P_\r+P_\nu=\frac {c^2}{3}\rho_{\r\nu},\,\, \Omega_{\r\nu}=\frac {\rho_{\r\nu}}{\rho_\cc}.
\ee Using introduced quantities, the second equation for the scale factor (\ref{eq:eqsscalfact}) is
rewritten in the form: \be \label{eq:HH0sqrtOmeg0a} H=\frac {\dot {a}}{a}=\frac {H_0}{a^2}\sqrt
{\Omega_{\r\nu}^0+\Omega_\d^0a+ \Omega_\Lambda^0a^4-\frac {kc^2}{R_0^2H_0^2}a^2}. \ee The first
equation has already been taken into account in formulas (\ref{eq:rhoaOmeg}). A solution of equation
(\ref{eq:HH0sqrtOmeg0a}) represents an implicit dependency of the scaling factor on time. At the
right hand of (\ref{eq:HH0sqrtOmeg0a}) under the square-root there is a fourth-order polynomial with
respect to $a$. This form of the solution was obtained by Lema\^{i}tre [\cite{Lemaitre1}]. Friedmann’s
solutions \cite{Fridmanpos, Fridmanneg} did not take into account the radiation, so that the
polynomial under the root was of the third order.

\subsection{Radiation, horizon, and distances} \label{sec:rad}

The equation of motion of a photon along the ray $\theta=\theta_0$, $\ff=\ff_0$ toward us follows
from the equality $\d s=0$, and connects its spatial and temporal coordinates: $\chi=\eta_0-\eta$.
At the instant of emission $\chi_\e=\eta_0-\eta_\e$. Since $\eta_\e\geq 0$, it follows that
$\chi_\e\leq\eta_0$. The equality $\chi_\e=\eta_0$ defines a spherical horizon; photons left this
horizon at the initial moment. For $\chi_\e>\eta_0$ a photon, even released in our direction, still
has not managed to reach us. This is a geometric horizon. There is also a physical horizon, which is
the sphere of the last scattering during cosmological recombination. One can look beyond it: the
theory of nucleosynthesis and the interpretation of distortions of the cosmic microwave background
or relic radiation allow us to do so. However, it is impossible in principle to look behind the
geometric horizon.

In cosmology, several concepts of distances can be introduced. McCrea \cite{McCrea} was the first to
pay attention to this, and gave definitions of the distances.

\begin{enumerate}

\item A metric distance $l$, which is the distance along the line of sight drawn from the observer
with fixed angles (see formula (\ref{eq:lmetrRetar})).

Other distances are determined by a common principle: the expression for any value in an expanding
and, generally speaking, non-planar space, is written down, and then this expression is equated to
the expression that would be true for the usual Euclidean space at a given distance. The distance is
named according to the quantity for which formulas are written. Usually, the following distances
from the observer are used (we define these at an arbitrary epoch, $t=t(\eta)$, but for the selected
point where we, humanity, are located).

\medskip

\item For the angular size. For the angular size. Let two signals issue at moment $t_{\rm source}=t(\eta-\chi)$ 
from two points located from the observer at equal distance
corresponding to the spatial coordinate $\chi$, and separated by an infinitesimally 
small angular distance, $\d\omega$. Let these signals arrive at the observer at moment $t=t(\eta)$. 
Then the linear distance between the points is 
$\d D_\ad=R(\eta-\chi)\sn_k(\chi)\d\omega=l_\ad\d\omega$. This implies that 
the distance $l_\ad=R(\eta-\chi)\sn_k\chi$ is the radius of the sphere with
quasi-center coinciding with the observer. The moment $\eta-\chi$ corresponds to
points for which this distance is determined. The distance becomes zero for $\chi=0$ (at the point of the 
observer, as one might
expect) and when $\chi=\eta$ (on the horizon). At some point the angular size has a minimum value.
This means that the angular size of objects with the same linear size decreases in the beginning as
the object recedes from the observer, and after passing the distance at which the angular size
reaches its minimum value, it increases. This is due to the fact that in remote areas that
correspond to earlier stages of expansion, the universe had a smaller scale, so that the lines of
sight were closer to each other. Similarly, when a rail of a certain size is crossing from one pole
of the Earth to another, first its angular size decreases, and then increases, since the meridians
converge approaching the poles.

\medskip

\item For the parallax $l_\pl=R(\eta)\sn_k\chi$, which is the radius of the sphere, but rather with
a quasicenter at the point to which the distance is measured, and at the time of the measurement.

\medskip

\item For the number of photons received by the observer from the source, taking into account the
difference in passage of time at the source and the observer,\\
$l_\nb=l_\pl\sqrt{R(\eta)/R(\eta-\chi)}$.

\medskip

\item For the apparent bolometric luminosity (called also photometric distance) $l_\bb\!=\!l_\pl
R(\eta)/R(\eta\!-\!\chi)$, where in addition to the difference in the passage of time, the loss of
radiative energy due to redshift is taken into account. If the luminosity of an object located at
the position corresponding to the radius of curvature $R(\eta)$ is equal to $L_\O$, then the
observed luminosity according to the definition of distance derived from the bolometric brightness
is equal to: \be \label{eq:Lbb} L_\bb=\frac {L_\O}{4\pi l_\bb^2}. \ee

\end{enumerate}

To obtain the current values of these distances, it is necessary to substitute $\eta=\eta_0$, and
$R(\eta_0)=R_0$. Modern distances are related as follows: \be \label{eq:lbbnbplad}
l_\bb^0=l_\nb^0\sqrt {1+z}=l_\pl^0(1+z)=l_\ad^0(1+z)^2=R_0\sn_k(\chi)(1+z). \ee Since $z\geq 0$, in
this chain of equalities the magnitude of the distances decreases from left to right.

The velocity of change of metric distance, which is the expansion rate at an arbitrary epoch,
$\eta$, complies with the Hubble--Lema\^{i}tre law: \be \label{eq:Hubblelaw} l=R(\eta)\chi,\quad
v=\dot {l}=\dot{R}\chi=\frac {\dot{R}}{R}l=Hl. \ee The Hubble distance, at which the expansion
velocity is equal to the speed of light, is $l_\H=c/H$; the current Hubble distance is
$l_\H^0=c/H_0$.

The relationship between speed and redshift is more complex than that between speed and distance
\cite{Harris}. At the current epoch the relation is: \be \label{eq:velocityredsht} \frac
{v}{c}=H_0\intl_0^z\frac {\d z}{H}. \ee This connection is model dependent and admits velocities
greater than the speed of light, so that the cosmological redshift is not identical to the classical
Doppler effect. The reason is that a photon changes its frequency not only at the instant of
emission from a moving source, which is taken into account by the Doppler effect, but experiences a
decrease in energy at each point of its flight to the observer due to the expansion of space, which
occurs according to the appropriate model. The expansion occurs identically with respect to any
point considered as a center. The existence of cosmological velocities higher than the speed of
light does not contradict the theory of relativity, since the mutual receding of points occurs not
because of their movement, but because of the expansion of space, across the complete span of which
no signals are transmitted.

\section{The Standard model ($\Lambda$CDM)}

\subsection{Model Parameters}

Modern cosmology has become a science based on observational data, which now have sufficient
accuracy to construct a model that adequately describes the real universe. The most important aspect
is the inference that space is very close to flat, which leads one to assume that $k=0$. In this
case the radius of curvature is infinitely large and should not appear in expressions for quantities
that have physical meaning. Therefore, as is often done, we adopt for its contemporary value the
Hubble distance: $R_0=l_\H^0=c/H_0$. Then the metric (\ref{eq:FRWmetr}) can be rewritten as: \be
\label{eq:metrplanespace} \d s^2=(l_\H^0)^2a^2(\eta)\left[\d\eta^2-\d\chi^2-\chi^2\d\omega^2\right].
\ee Of all the cosmological parameters, the current temperature of the radiation, which is very
close to thermal (deviations from a blackbody spectrum are of order $10^{-5}\div 10^{-4}$) and
called the cosmic microwave or relict background, has been determined with the greatest in cosmology
accuracy: its value is $T_0=2.7277$ K. At an arbitrary epoch corresponding to redshift $z$,
$T=T_0/a=T_0(1+z)$. The temperature of the neutrinos is connected to that of the radiation as
$T_\nu=\sqrt[3]{4/11}\,T=0.71377\,T$, $T_\nu^0=1.9469$ K. The coefficient is obtained from the
consideration that, due to the adiabatic expansion, the entropy of the total mixture of matter and
radiation does not change, while during the annihilation of electron-positron pairs their entropy
passes to the radiation \cite{AlphHerm}. The entropy of the neutrino gas depends only on its
temperature, and does not change.

Since radiation and neutrinos are ultrarelativistic, their mass densities are proportional to the
fourth power of their temperature. For radiation according to the Stefan-Boltzmann formula, \be
\label{eq:rhor0} \rho_\r^0=\frac {\aSB}{c^2}T_0^4=4.66\cdot 10^{-34}\,\,\hbox{{\rm g/cm}}^3, \ee
where $\aSB=(8\pi^5h/15c^3)(\kB/h)^4$ is called the Stefan constant. For six types of neutrinos,
which are fermions rather than bosons, \be \label{eq:rhonu0} \rho_\nu^0=6\cdot\frac {7}{8}\cdot\frac
{\aSB}{c^2}(T_\nu^0)^4=6.35\cdot 10^{-34}\,\,\hbox{{\rm g/cm}}^3. \ee Together, radiation and
neutrinos have a density \be \label{eq:rhornu0} \rho_{\r\nu}^0=1.10\cdot 10^{-33}\,\,\hbox{{\rm
g/cm}}^3. \ee The Hubble constant, according to the latest definitions, is known to within several
percent: $H=70\pm 3$ km/s/Mpc \cite{WMAPparameters,Planck18112376}. Here we adopt
$H_0=70\,\hbox{{\rm km/s/Mpc}}=2.27\cdot 10^{-18}\hbox{{\rm 1/s}}$, so that the current critical
density and the Hubble distance are equal to: \beq \label{eq:rhoc0stand} & \strut\disp
\rho_\cc^0=\frac {3H_0^2}{8\pi G}=9.207\cdot 10^{-30}\, \hbox {{\rm g/cm}}^3, & \\
\label{eq:lH0stand} & \strut\disp l_\H^0=\frac {c}{H_0}=1.3215\cdot 10^{28}\,\,\hbox {{\rm cm}}=
14.2\,\,\hbox {{\rm G light yrs}}=4.2828\,\,\hbox {{\rm Gpc}}. & \nonumber \eeq Current relative
fractions of radiation, neutrinos, and their sum are obtained as follows: \be \label{eq:Omegarnu0}
\Omega_\r^0=5.06\cdot 10^{-5},\quad\Omega_\nu^0=6.90\cdot 10^{-5},\quad \Omega_{\r\nu}^0=1.196\cdot
10^{-4}. \ee The main gravitational component of the mass of the universe, according to modern
concepts, is dark energy; its share is estimated as $0.721\pm 0.035$ \cite{WMAPparameters}. Let's
take the value $\Omega_\Lambda^0=0.72$, so that $\rho_\Lambda^0=6.63\cdot 10^{-30}$ g/cm$^3$. Since
space is flat, $\rho_\t=\rho_\cc$ and $\Omega_\t=\Omega_\t^0=1$. The rest is a fraction of the dust,
$\Omega_\d^0=1-\Omega_{\r\nu}^0-\Omega_\Lambda^0=0.27988\approx 0.28$ and $\rho_\d^0=2.577\cdot
10^{-30}$ g/cm$^3$.

Cosmological densities are very low, much lower than current densities in astronomical objects. Even
in interstellar space, in each cubic centimeter there is an average of $\sim 1$ hydrogen atom. The
densities of the cosmological components correspond to the following numbers of hydrogen atoms in a
cubic meter (not cm): 
\be \label{eq:hydratoms}
10^6\frac{\rho_\cc^0}{m_\H}\!=\!5.5,\,10^6\frac{\rho_\Lambda^0}{m_\H}\!=\!4.0,
\,10^6\frac{\rho_\d^0}{m_\H}\!=\!1.5,\,10^6\frac{\rho_\r^0}{m_\H}\!=\!2.8\!\cdot\!10^{-4},\nonumber\ee
\be 10^6\frac{\rho_\nu^0}{m_\H}\!=\!3.8\!\cdot\!10^{-4},\,
10^6\frac{\rho_{\r\nu}^0}{m_\H}\!=\!6.6\!\cdot\!10^{-4}, \ee
where $m_\H=1.67\cdot 10^{-24}$ g. At
the same time $1$ cm$^3$ contains $n_\ph^0=20.286T_0^3=412$ relict photons and $\disp
6\cdot\frac{3}{4}\cdot \frac{4}{11}n_\ph^0=674$ relict neutrinos. Using the density of dark energy,
the current value of the cosmological constant is determined as
$\disp\Lambda=3\frac{\Omega_\Lambda^0}{(l_\H^0)^2}=1.24\cdot 10^{-56}$\,\, cm$^{-2}$. Note that
during inflation this density was equal to the Planck density: $\disp\rho_\Lambda=\frac{\Lambda
c^2}{8\pi G}=\rho_\Pl=\frac{c^5}{G^2\hbar}= 5.1593\cdot 10^{92}$ g/cm$^3$, so then it was
$\Lambda_\Pl=9.6\cdot 10^{66}$ \,cm$^{-2}$.

\subsection{Basic dependencies}

Substituting into equation (\ref{eq:HH0sqrtOmeg0a}) $k=0$ and dividing the variables, we obtain the
relationship between the time and scale factors. Using the relationship $c\d t=l_\H^0a(\eta)\d\eta$,
and the relationship between the time coordinate and $a$, we derive: \be \label{eq:eqaOmegtzetpl}
\intl_0^a\frac {a\d a}{\sqrt {\Omega_{\r\nu}^0+\Omega_\d^0a+
\Omega_\Lambda^0a^4}}=H_0t,\quad\intl_0^a\frac {\d a} {\sqrt
{\Omega_{\r\nu}^0+\Omega_\d^0a+\Omega_\Lambda^0a^4}}=\eta. \ee If we introduce the notation
($\Omega_{\r\nu}^0+\Omega_\d^0+\Omega_\Lambda^0=\Omega_\t=1$), 
\be \label{eq:xbetamem}
H_\Lambda=H_0\sqrt {\Omega_\Lambda^0},\quad x_0=\left(\frac
{\Omega_\Lambda^0}{\Omega_{\r\nu}^0}\right)^{1/4},\nonumber\ee
\be \beta=\frac
{\Omega_\d^0}{(\Omega_{\r\nu}^0)^{3/4}(\Omega_\Lambda^0)^{1/4}}
\eta_*=\left(\Omega_{\r\nu}^0\Omega_\Lambda^0\right)^{-1/4}, \ee 
and make the change of
variable $a=x/x_0$, then the equation (\ref{eq:HH0sqrtOmeg0a}) will be transformed into \be
\label{eq:Hofx} H=\frac {\dot {x}}{x}=H_\Lambda\frac {\sqrt {1+\beta x+x^4}}{x^2}, \ee and the
relations between the variables take the form \be\label{eq:H0tI1xbeta} H_\Lambda
t=I_1(x,\beta),\quad\eta=\eta_*I_0(x,\beta),\quad I_j(x,\beta)=\intl_0^x\frac {x^j\d x}{\sqrt
{1+\beta x+x^4}}. \ee The parameter of the integrals with variable upper limit is
$\beta\!=\!265.69$. The values of the constants $H_\Lambda\!=\!59.397\,\,\hbox {{\rm
km/s/Mpc}}\!=1.9249\cdot 10^{-18}$s$^{-1}$, $\disp x_0=8.8088$, $\eta_*=10.381$. The age of the
universe according to the Standard model with the adopted values of the parameters is
$t_0=I_1(x_0,\beta)/H_\Lambda=4.33\cdot 10^{17}$s$=13.722$~Gyr.

The two integrals are computed numerically, although approximate representations of the integrals
are possible as well. For small $x$, relatively simple formulas can be obtained:
\begin{widetext}
\beq\label{eq:I0aprsmallx} & \strut\disp  \hspace {-12pt} I_0(x,\beta)\sim 2\frac {x}{q}\!-\! \frac
{1}{35}\frac {x^5}{r}\frac {1}{q^2}\left(5+\frac {10}{q}+\frac {12}{q^2}+\frac
{8}{q^3}\right)+ & \nonumber \\ & \strut\disp +\frac
{x^9}{1716r^3q^2} \left(99+\frac {198}{q}+\frac
{252}{q^2}+\frac {216}{q}+\frac {80}{q^4}-\frac {96}{q^5}-\frac {192}{q^6}-
\frac {128}{q^7}\right), & \\ \label{eq:I1aprsmallx} & \strut\disp  
I_1(x,\beta)\sim\frac
{2}{3}\frac {q+1}{q^2}x^2- \frac {x^6}{63rq^2}\left(7+\frac {14}{q}+\frac {18}{q^2}+\frac {16}{q^3}+
\frac {8}{q^4}\right)+ & \nonumber \\ & \strut\disp +\frac {x^{10}}{2860r^3q^2}\left(143+\frac
{286}{q}+ \frac {374}{q^2}+\frac {352}{q^3}+\frac {200}{q^4}-\frac {32}{q^5}- \frac {224}{q^6}-\frac
{256}{q^7}-\frac {128}{q^8}\right). & \eeq 
\end{widetext}
Here $r=\sqrt{1+\beta x}$, $q=1+r$. These formulas
represent the integral $I_1$ with a relative discrepancy of $10^{-6}$ for $x\leq 1.9$, $10^{-5}$ for
$x\leq 2.5$, and $10^{-4}$ for $x\leq 3.2$. The accuracy of the formula for $I_0$ is somewhat
higher: the value of $10^{-6}$ is already achieved for $x\leq 2.1$, $10^{-5}$ for $x\leq 2.9$, and
$10^{-4}$ for $x\leq 3.6$.

For large values of the argument, the behavior of the integrals is substantially different. The
integral $I_0$ from $x\to\infty$ has a finite limit, while $I_1$ tends to infinity. Approximately,
they can be represented as follows: 
\begin{widetext}\beq \label{eq:I0aprlrgx} & \strut\disp I_0(x,\beta)\sim
I_0(\infty,\beta)-\frac {1}{x} \left(1+\frac {\beta}{x^3}\right)^{1/2}F\left(1,\frac {5}{6},\frac
{4}{3},- \frac {\beta}{x^3}\right), & \nonumber \\ & \strut\disp
I_1(x,\beta)\!\sim\!\ln x+S_0(\beta/x^3)\!+\!P(x_*,\beta),\,\, &
\\ \label{eq:I1aprlrgx} & \strut\disp P(x_*,\beta)\!=\!I_1(x_*,\beta)\!-\!\ln
x_*\!-\!S_0(\beta/x_*^3),\,\, S_0(u)\!=\!\frac {1}{3}\sum_{n=1}^\infty\frac
{(2n-1)!!}{n(2n)!!}(-u)^n.~~ & \eeq\end{widetext} 
Here $F(a,b,c,x)$ is the hypergeometric function. For $x_*$, we
can take the value of $10$. For $\beta=265.69$, the values of the integrals in the last formulas
are: $I_0(\infty,\beta)=0.42880$, $I_1(10,\beta)=0.94380$, and $P(10,\beta)=-1.3992$. Calculations
using formula (\ref{eq:I0aprlrgx}) give the value of $I_0(x,\beta)$ with five significant digits
when $x\geq 7.3$, and with (\ref{eq:I1aprlrgx}) five significant digits of $I_1(x,\beta)$ are
obtained when $x\geq 8.5$.

It should be emphasized that the scale factor $a$ and the redshift $z$ are tied to the current
epoch, and that they change with increasing age of the universe. At the same time, the variable $x$
is associated only with time $t$ (through the radius of the curvature $R$), while the parameters
$\beta$, $H_\Lambda$, and $\eta_*$ are strictly constant. Indeed, products $\disp
M_\d=\frac{4\pi}{3}\rho_\d R^3=2.49\cdot 10^{55}$~g (the mass of dust matter in a sphere of radius
$R$) and $W=4\pi\rho_{\r\nu}R^4=4.22\cdot 10^{80}$~g $\cdot$~cm do not depend on time, as is also
the case for the density $\rho_\Lambda=\rho_\Lambda^0$, which is proportional to the cosmological
constant $\Lambda$. These values can be used to express the variable $x$ and other parameters: 
\be
\label{eq:txbetaetast} x\!=\!\left(\!\frac {4\pi\rho_\Lambda}{W}\!\right)^{1/4}R,\,\,
H_\Lambda\!=\!\sqrt {\frac {\Lambda}{3}}c,\,\, \beta\!=\!\frac
{3M_\d}{W^{3/4}(4\pi\rho_\Lambda)^{1/4}}, \nonumber\ee
\vspace{-0.5cm}
\be\eta_*\!=\!\left(\!\frac {9}{16\pi}\frac
{c^4}{G^2W\rho_\Lambda}\!\right)^{1/4} \!=\!\left(\!\frac {9}{2}\frac
{c^2}{GW\Lambda}\!\right)^{1/4}. \ee
The variable $\eta$ is directly connected with time and
expressed through $\eta_*$ and $x$.

\subsection{Roles of components at different epochs}

In the expressions for total mass density \be \label{eq:rhocx}
\rho_\t=\rho_\cc=\rho_\d+\rho_{\r\nu}+\rho_\Lambda=\rho_\cc^0\Omega_\Lambda^0 \frac {1+\beta
x+x^4}{x^4} \ee and gravitational mass density \be \label{eq:rhogx}
\rho_\g=\rho_\d+2\rho_{\r\nu}-2\rho_\Lambda= \rho_\cc^0\Omega_\Lambda^0\frac {2+\beta x-2x^4}{x^4}
\ee the mass density of dark energy is constant, while others decrease with time. Therefore, at
different epochs the components have played different roles.

At certain points in time, the densities become equal. Since the components give different
contributions to the gravitational mass density --- the radiation gives a double positive
contribution, and the vacuum gives a double negative one --- their effect on the gravitation is
different at different times. All of these moments are given in Table 1, which lists the values of
the parameter $x$, the redshift $z$, and the coordinate $\eta$, the fraction of the full age and the
age of the universe itself at the corresponding moments, as well as the time elapsed from these
moments to the present epoch. The gravitational mass density becomes zero at a value of $x$
determined by the equation $x^4-(\beta/2)x-1=0$. Moments when $\rho_\d=\rho_\Lambda$ and when
$\rho_\g=0$ almost coincide, because the radiation and neutrino densities are small at these
moments. The moment when $\rho_{\r\nu}=\rho_\Lambda$ corresponds to the time when $x$ is very close
to $1$. 
\begin{table*}[tbp]
\centering 
\caption{\label{tab:1} Epochs of equality of densities and forces.}
\smallskip\vspace{3mm} 
\begin{tabular}{lllllll}
\hline
Epoch & $x$ & $z$ & $\eta$ & $t/t_0$ & $t$ Gyrs & $t_0-t$ \\ \hline $\rho_\d=\rho_\r$ &
$0.00159$ & $5529$ & $0.0151$ & $1.34\cdot 10^{-6}$ & $1.85\cdot 10^{-5}$ & $13,7$ \\
$\rho_\d=\rho_\nu$ & $0.00217$ & $4057$ & $0.0200$ & $2.41\cdot 10^{-6}$ & $3.31\cdot 10^{-5}$ &
$13.7$ \\ $\rho_\d=2\rho_\r$ & $0.00319$ & $2764$ & $0.280$ & $4.90\cdot 10^{-6}$ & $6.72\cdot
10^{-5}$ & $13.7$ \\ $\rho_\d=\rho_{\r\nu}$ & $0.00376$  & $2339$ & $0.0324$ & $6.64\cdot 10^{-6}$ &
$9.11\cdot 10^{-5}$ & $13.7$ \\ $\rho_\d=2\rho_\nu$ & $0.00434$  & $2028$ & $0.365$ & $8.59\cdot
10^{-6}$ & $1.72\cdot 10^{-4}$ & $13.7$ \\ $\rho_\d=2\rho_{\r\nu}$ & $0.00752$  & $1169$ & $0.0572$
& $2.27\cdot 10^{-5}$ & $3.11\cdot 10^{-4}$ & $13.7$ \\ $\rho_{\r\nu}=\rho_\Lambda$ & $1.0000$  &
$7.809$ & $1.198$ & $0.0488$ & $0.669$ & $13.0$ \\ $\rho_\d=2\rho_\Lambda$ & $5.1025$  & $0.7264$ &
$2.7138$ & $0.5261$ & $7.219$ & $6.5$ \\ $\rho_\g=0$ & $5.1050$ & $0.7255$ & $2.7144$ & $0.5264$ &
$7.224$ & $6.5$ \\ $\rho_\d=\rho_\Lambda$ & $6.4288$ & $0.3702$ & $2.983$ & $0.7043$ & $9.66$ &
$4.06$ \\ Current & $8.8088$ & $0$ & $3.32$ & $1$ & $13.7$ & $0$ \\
\hline
\end{tabular}
\end{table*}

\subsection{Distances, speeds, acceleration: past, current, and future}

In the flat model, $\sn_0(\chi)=\chi$, the quasicenter and real center of spheres coincide, the
parallax distance and the radius of a sphere are equal to the metric distance: $l_\pl=r=l$. In the
Standard model the expressions for $l/l_H^0$ and the dimensionless velocity of the expansion $v/c$
coincide as well. Indeed, at any moment: \be \label{eq:vdctl} \frac {v}{c}=\frac {\dot {l}}{c}=\frac
{H}{c}l=\frac {l}{l_H}. \ee In the Standard model the metric distance from the observer in the
current universe to a location with coordinate $\chi$ is given by the formula following from
(\ref{eq:Hubblelaw}) and (\ref{eq:H0tI1xbeta}): \be \label{eq:metrdiststand}
l^0=R(\eta_0)\chi=R_0a(\eta_0)\chi=l_\H^0(\eta_0-\eta)=l_\H^0\eta_* [I_0(x_0,\beta)-I_0(x,\beta)].
\ee The equalities (\ref{eq:lbbnbplad}) can be rewritten as \be \label{eq:lbbnbpladstmod}
l_\bb^0=l_\nb^0\sqrt {1+z}=l_\pl^0(1+z)=l_\ad^0(1+z)^2=R_0\chi(1+z)=l^0(1+z). \ee In what follows,
we refer mainly to current values and use dimensionless distances, measuring them in terms of the
Hubble distance according to the scheme $\tl=l/l_\H^0$. Therefore, all distances are expressed (as
$v/c$ in (\ref{eq:vdctl})) via the metric distance: 
\be \label{eq:distatsnd} \tl_\pl=\frac
{v}{c}=\tl=\eta_*[I_0(x_0,\beta)-I_0(x,\beta)],\quad \tl_\ad=\tl a,\nonumber\ee
\vspace{-1cm}\be\tl_\nb=\frac {\tl}{\sqrt{a}},\quad\tl_\bb=\frac {\tl}{a}, 
\quad a=\frac {x}{x_0}=\frac {1}{1+z}. \ee 

\begin{figure*}[tbp]
\centering
\includegraphics[width=12 cm]{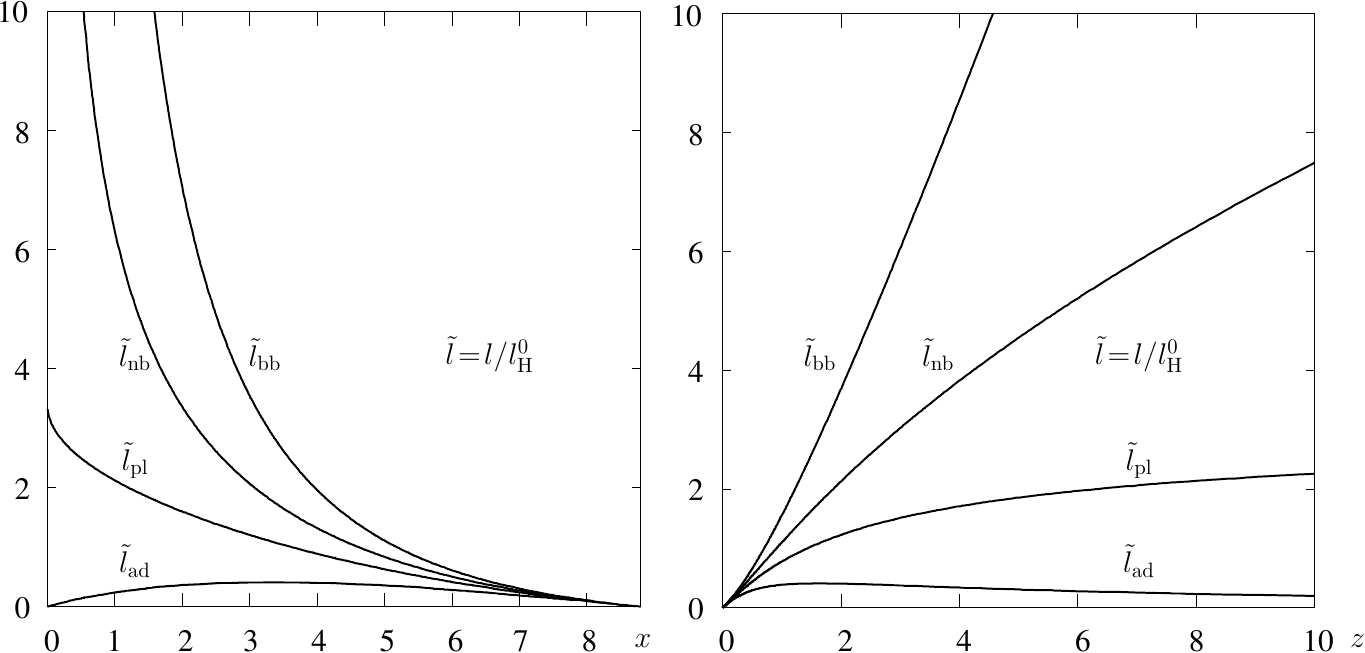} \caption {\label{fig:dlxz} Distances as functions of $x$
({\it left}) and $z$ ({\it right}).} 
\end{figure*} Figure~\ref{fig:dlxz} plots dependences of
distances on the variable $x$ ({\it left}) and redshift $z$ ({\it right}).

Let us consider three additional moments in time corresponding to particular events. The first event
was the phisical horizon (about $z=1000$), the second was when the angular size distance had its
maximum value ($z=1.6302$), and the third was when the current metric distance equaled the Hubble
distance ($z=1.4233$). These data, and also for comparison, the moments corresponding to several
characteristic values of redshift, are given in Table 2, serving as a continuation of Table 1.

\begin{table*}[tbp] 
\centering \caption {\label{tab:2} Epochs associated with the characteristic
values of redshift.} \smallskip\vspace{3mm} 
\begin{tabular}{lllllll}
\hline
 $x$ & $z$ & $\eta$ & $t/t_0$ & $t$Gyrs & $t_0-t$ \\ \hline 
 $0.0088000$ & $1000$ & $0.064629$ & $2.9659\cdot 10^{-5}$ & $4.0697\cdot 10^{-4}$ & $13.721$ \\ 
 $0.017582$ & $500$ & $0.10796$ & $9.4743\cdot 10^{-5}$ & $1.3000\cdot 10^{-3}$ & $13.720$ \\ 
 $0.087215$ & $100$ & $0.30606$ & $1.2021\cdot 10^{-3}$ & $0.016494$ &$13.705$ \\ 
 $0.17272$ & $50$ & $0.45696$ & $3.4279\cdot 10^{-3}$ & $0.047036$ & $13.675$ \\
$0.80080$ & $10$ & $1.0642$ & $0.034925$ & $0.47924$ & $13.242$ \\ 
$0.97875$ & $8$ & $1.1841$ &$0.047233$ & $0.64811$ & $13.074$ \\ 
$3.3491$ & $1.6302$ & $2.2320$ & $0.29360$ & $4.0288$ &$9.6930$ \\ 
$3.6350$ & $1.4233$ & $2.3224$ & $0.33008$ & $4.5293$ & $9.1925$ \\ 
$2.2022$ & $3$ &$1.8084$ & $0.15891$ & $2.1806$ & $0.11541$ \\ 
\hline
\end{tabular}
\end{table*} 
Table 3 lists the values of the distances to the points indicated in Tables 1 and 2. 
\begin{table*}[tbp] \centering
\caption {\label{tab:3} Distances to characteristic points of the Standard Model.}
\smallskip\vspace{3mm}\begin {tabular}{lllllll} 
\hline
$z$ &$\tl$ & $\tl_\ad$ & $\tl_\nb$ & $\tl_\bb\phantom{\Bigl|}$ \\ \hline $\infty$ & $3.322$  & $0$       
          & $\infty$ & $\infty$ \\ $5529$   & $3.307$  & $5.981\cdot 10^{-4}$ & $245.9$  & $18288$
\\ $4057$   & $3.302$  & $8.138\cdot 10^{-4}$ & $210.4$  & $13401$ \\ $2764$   & $3.294$  &
$1.192\cdot 10^{-3}$ & $173.2$  & $9108$ \\ $2339$   & $3.290$  & $1.406\cdot 10^{-3}$ & $159.2$  &
$7700$ \\ $2028$   & $3.286$  & $1.619\cdot 10^{-3}$ & $148.0$  & $6667$ \\ $1169$   & $3.265$  &
$2.790\cdot 10^{-3}$ & $111.7$  & $3821$ \\ $1000$   & $3.258$  & $3.255\cdot 10^{-3}$ & $103.1$  &
$3261$ \\ $500$    & $3.214$  & $6.416\cdot 10^{-3}$ & $71.95$  & $1610$ \\ $100$    & $3.016$  &
$0.02987$            & $30.31$  & $304.7$ \\ $50$     & $2.865$  & $0.05619$            & $20.46$  &
$146.1$ \\ $10$     & $2.258$  & $0.2053$             & $7.490$  & $24.83$ \\ $8$      & $2.138$  &
$0.2376$             & $6.415$  & $19.25$ \\ $7.809$  & $2.125$  & $0.2412$             & $6.306$  &
$18.72$ \\ $3$      & $1.514$  & $0.3785$             & $3.028$  & $6.056$ \\ $1.630$  & $1.090$  &
$0.4146$             & $1.768$  & $2.868$ \\ $1.423$  & $1.000$  & $0.4126$             & $1.557$  &
$2.423$ \\ $0.7264$ & $0.6086$ & $0.3525$             & $0.7996$ & $1.051$ \\ $0.7255$ & $0.6100$ &
$0.3524$             & $0.7987$ & $1.049$ \\ $0.3702$ & $0.3397$ & $0.2479$             & $0.3977$ &
$0.4655$ \\ $0$      & $0$      & $0$                  & $0$      & $0$ \\ 
\hline
\end {tabular}
\end{table*} The rate of change of the parallactic distance coincides with $v/c$, since this distance
corresponds to the metric distance. The rates of recession of the remaining distances are determined
from their definitions by differentiation with respect to time while keeping the spatial coordinate
$\chi$ fixed. The expressions for these velocities are given in Table 4.

\begin{table*}[tbp] \centering \caption{\label{tab:4} Current velocities of recession at different
distances.} \smallskip\vspace{3mm}\begin {tabular}{ccccc}\hline
Metric & ad & pl & nb & bb \\ \hline $H_0l$ & $H(\eta_0\!-\!\chi)l_\ad$ & $H_0l_\pl$ &
$\disp\frac{3H_0\!-\!H(\eta_0\!-\!\chi)}{\,\,\,\,\,\,\,\,2\phantom{bigl|}} l_\nb$ &
$[2H_0\!-\!H(\eta_0\!-\!\chi)]l_\bb$ \\ \hline \end {tabular}\end{table*} The acceleration of the
cosmological expansion is determined by the first equation in (\ref{eq:eqsscalfact}): \be
\label{eq:acceleratl} \dot {v}=\ddot {l}=\DR {^2}{t^2}l_\H^0a\chi=l_\H^0\ddot{a}\chi=
\frac{\ddot{a}}{a}l=-\frac{4\pi G}{3}\rho_\g l=H_\Lambda^2 \frac {x^4-\beta x/2-1}{x^4}l. \ee As
already mentioned, in the gravitational mass density $\rho_\g=\rho_\d+2\rho_\r-2\rho_\Lambda$
densities $\rho_\d$ and $\rho_\r$ decrease with increasing age of the universe, while
$\rho_\Lambda=\rho_\Lambda^0$. Therefore, in the numerator of the last fraction in
(\ref{eq:acceleratl}), the relative importance of the first term increases with time. At the present
time ($x=x_0$), the gravitational mass density is negative:
$\rho_\g^0=\rho_\d^0+2\rho_\r^0-2\rho_\Lambda^0=-1.0677\cdot 10^{-29} \,\,\hbox {{\rm g/cm}}^3$, so
that the expansion occurs with an acceleration. But the acceleration at the current Hubble distance
(speed equal to the speed of light) is only
\be \label{eq:dotv0} \dot {v}_\H^0=-\frac {4\pi
G}{3}\rho_\g^0l_\H^0=\frac {H_0c}{2}(2 \Omega_\Lambda^0-\Omega_\d^0-2\Omega_{\r\nu}^0)=\nonumber\ee
\be 3.94\cdot
10^{-8}\,\, \hbox {{\rm cm/s}}^2\approx 4\,\,\Angs\,/\hbox {{\rm s}}^2. \ee In the distant future at
$t\to\infty$ ($\eta_\infty=4.4514$) 
\be \label{eq:farfuture} a=\frac {1}{1+z}\sim\left(\frac
{\Omega_\d^0}{4\Omega_\Lambda^0}\right)^{1/3} e^{H_\Lambda t}=0.46000\,e^{H_\Lambda t},\nonumber\ee
\vspace{-0.5cm}\be x\sim 4.0520\,e^{H_\Lambda t},\quad \eta\sim\eta_\infty-2.5619e^{-H_\Lambda t}. \ee 
Thus, the scale of the
universe will increase exponentially, so that a second inflation will take place, which we will
discuss in more detail later. However, according to (\ref{eq:farfuture}), an exponential expansion
will really begin only at $t\sim t_\Lambda=1/H_\Lambda$. The time scale is $1/H_0=4.4081\cdot
10^{17}$s$=13.969$~Gyr, $t_\Lambda=1/H_\Lambda=5.1950\cdot 10^{17}$~s$=16.462$~Gyr. We also define
the distance $l_\Lambda=c/H_\Lambda=l_\H^0/\sqrt{\Omega_\Lambda^0}=1.5574\cdot
10^{28}$~cm$=5.0473$~Gpc.

The speed of expansion of space at the Hubble distance is, by definition, equal to the speed of
light. The velocity of recession of the Hubble distance is derived using equation
(\ref{eq:dotHrho}): \be \label{eq:dotlH} \dot {l}_\H=\DR {}{t}\frac {c}{H}=-\frac {c}{H^2}\dot
{H}=\frac {c}{H^2} \left(H^2+\frac {4\pi G}{3}\rho_\g\right)=\nonumber \ee
\be c\left(1+\frac {1}{2} \frac
{\rho_\g}{\rho_\cc}\right)=\frac {c}{2}\frac {4+3\beta x}{1+\beta x+x^4}. \ee According to this
formula, at the beginning of the expansion the velocity is close to the two speeds of light,
decreasing with time, and in the distant future it will approach zero. Acceleration at the Hubble
distance increases with time, but remains finite: \be \label{eq:accelerstar} \dot {v}_\H=-\frac
{4\pi G}{3}\rho_\g l_\H=-\frac {4\pi G}{3}\rho_\g \frac {c}{H}=H_\Lambda c\frac {x^4-\beta
x/2-1}{x^2\sqrt {1+\beta x+x^4}}\to\nonumber \ee
\be  H_\Lambda c=5.77\Angs\,/\hbox {{\rm s}}^2. \ee Acceleration of
the distance itself is negative: \be \label{eq:acclH} \ddot {l}_\H=-\frac {c}{2}\frac
{H_\Lambda}{x}\frac {\beta+16x^3+9\beta x^4} {(1+\beta x+x^4)^{3/2}}\sim -cH_\Lambda\frac
{9}{2}\frac {\beta}{x^3}. \ee From these formulas it is clear that the accelerations are of the same
order as the product of the speed of light and the current Hubble constant, or the asymptotic value
of the Hubble parameter: $cH_0=3\cdot 10^{10}\cdot 2.27\cdot 10^{-18}=6.81\cdot 10^{-8}$ cm/s$^2$,
$cH_\Lambda=5.77\cdot 10^{-8}$ cm/s$^2$ (which coincides with the limit of (\ref{eq:accelerstar})).

\subsection{Evolution of redshift and apparent luminosity}

As mentioned above, the scale factor $a$, and therefore the redshift $1+z=1/a$ are tied to the epoch
of observation. Therefore, the value of $z$ for each sufficiently distant object should change with
increasing age of the universe. Therefore, luminosities of objects should change as well. A.~Sandage
drew attention to this problem. He calculated the changes for the model of ``dust'' with different
values of $\Omega_\d^0$ (\cite{AlSand}). In the Appendix \cite {McVittie} to the paper
\cite{AlSand}, McVittie made the same calculations while adding a cosmological term. Later A.Loeb
\cite{Loeb} (apparently independently) proposed to determine changes in the redshift $z$ of quasars
using observations of the $L_\alpha$-forest with the 10-meter Keck telescope. He also transformed
these changes into changes of the velocities of radiating objects. Such changes are known as the
Sandage-Loeb effect.

Let us find the dependence of the change of $z$ on the age of the universe according to the Standard
model. Since for this relation the dependence of the radius of curvature on time is significant, we
will write $R(t)$ without changing the notation. The redshift of lines in the spectrum of some
object at a location corresponding to time $t=t(\eta)$ from the beginning of the expansion, and
observed at a position corresponding to the fixed time $t_0=t(\eta_0)$, is determined by the well
known formula $1+z=R(t_0)/R(t)$. Then $z$ is uniquely related to time $t$, and equal to 0 at the
observer's location, $z=0$. For the complete definition of $z$, both times should be specified as
arguments, i.e., $z(t,t_0)$. However, this is traditionally not done, since the epoch of $t_0$ is
fixed; in the past $z>0$ and in the future $-1<z<0$ with respect to $t_0$. At this point we adopt a
more detailed designation.

After some time has passed, the age of the universe has increased and the epoch to which redshifts
are attached has moved to the moment $t_0'=t(\eta_0')$. Then an object at a given redshift has moved
to time $t'=t(\eta')$ without changing its spatial coordinate $\chi$. A connection between the
moments of emission of radiation and its reception by the observer does not change in terms of the
conformal coordinates, and the difference between the times of the observer and the object is
preserved: \be \label{eq:eqphotont0st}
\chi=\eta_0'-\eta'=\eta_0-\eta,\quad\eta_0'-\eta_0=\eta'-\eta. \ee In particular, the infinitesimal
displacements are equal as well: $\d\eta_0=\d\eta$. Using the relation $c\d t=R(t)\d\eta$ at times
$t$ and $t_0$, we obtain the relation between the differentials of time and the derivative of one
with respect to the other: \be \label{eq:dtdt0} \d t_0=\frac {R(t_0)}{c}\d\eta_0=\frac
{R(t_0)}{c}\d\eta=\frac {R(t_0)}{c} \frac {c}{R(t)}\d t=\frac {R(t_0)}{R(t)}\d t,\nonumber\ee
\vspace{-0.5cm}\be \DR{t}{t_0}=\frac {R(t)}{R(t_0)}=\frac {1}{1+z}. \ee The last relation between the passage of time of
the object and the observer has already been used in section \ref{sec:rad} for the transformation
from a parallactic distance to a distance measured according to the flux of photons.

To detect changes of $z$, one must measure shifts of lines in the spectrum of a source (with the
same value of the $\chi$ coordinate) at different times. The difference between the times should be
much smaller than the times themselves, so increments of values can be replaced by their
differentials (infinitesimally small), and it is sufficient to determine the derivatives of the
variables. Using (\ref{eq:dtdt0}) we find: \be \DR {R(t)}{t_0}=\DR {R(t)}{t}\DR {t}{t_0}=\dot
{R}(t)\frac {R(t)}{R(t_0)},\quad \DR {R(t_0)}{t_0}=\dot {R}(t_0). \ee From the latter we obtain
(\cite{AlSand}) \be \label{eq:Dzdtz} \DR {z}{t_0}=\DR {(1+z)}{t_0}=\DR {}{t_0}\frac {R(t_0)}{R(t)}=
\frac {\dot {R}(t_0)}{R(t)}-\frac {R(t_0)}{R^2(t)}\dot {R}(t) \frac {R(t)}{R(t_0)}=\nonumber\ee
\vspace{-0.5cm}\be \frac {\dot
{R}(t_0)}{R(t_0)}\frac {R(t_0)}{R(t)}- \frac {\dot {R}(t)}{R(t)}=H_0(1+z)-H. \ee The dependence of
$H$ on $z$ is derived if $x$ is substituted by $x_0/(1 + z)$ in (\ref{eq:Hofx}).

A chaA change in the redshift will result in a change in the observed luminosity of objects. The rate 
of change of the photometric distance in the current epoch, as follows from the equalities
(\ref{eq:Lbb}), (\ref{eq:lbbnbpladstmod}) (its boundary parts $l_\bb^0=l^0(1+z)$) and
(\ref{eq:Dzdtz}), is equal (in accordance with Table 4) to:
\be \DR {l_\bb^0}{t_0}=\dot{l}^0(1+z)+l^0\DR {z}{t_0}=H_0l^0(1+z)+ l^0[H_0(1+z)-H]=\nonumber\ee
\vspace{-0.5cm}\be \left(2H_0-\frac {H}{1+z}\right)l_\bb^0. \ee
Then \be \dot {L}_\bb^0\!=\!-2\frac {L_\O}{4\pi(l_\bb^0)^3}\DR {l_\bb^0}{t_0}\!=\!- 2\frac
{L_\bb^0}{l_\bb^0}\DR {l_\bb^0}{t_0}\!=\!-2L_\bb^0 \left(\!2H_0-\frac {H}{1\!+\!z}\!\right)\!,
\nonumber\ee
\vspace{-0.5cm}\be \frac {1}{H_0}\DR {\ln L_\bb^0}{t_0}\!=\!-2\left(2\!-\!\frac {1}{1\!+\!z} \frac
{H}{H_0}\!\right)\!. \ee

\begin {figure}[ht] \centering \includegraphics[width=7 cm]{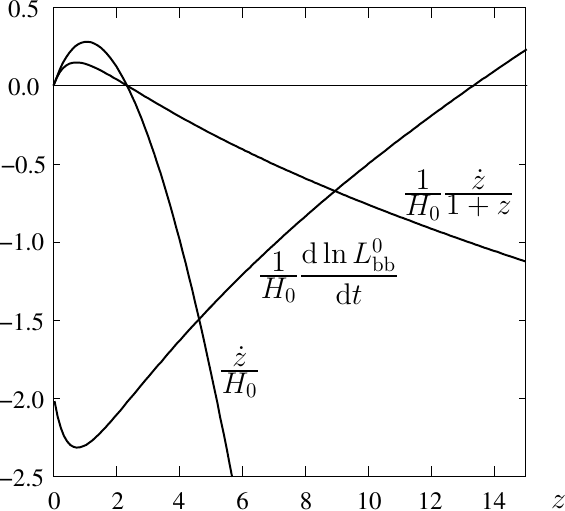}
\caption {\label{fig:tcz} Changes in redshift and luminosity as a function of $z$.} \end {figure}

Figure~\ref{fig:tcz} (for brevity, the derivative $\d z/\d t_0$ is denoted by $\dot{z}$) presents
the dependencies of $\dot{z}/H_0$ and ratio $\dot{z}/[H_0(1+z)]$ on the (current) redshift $z$.
First, the speed $\dot{z}$ is positive, i.e., $z$ increases; at $z=2.34$ the derivative $\dot{z}$
becomes zero. Between two zeros (at $z=0$ and $z=2.34$) at the point $z=1.06$ there is a maximum
equal to $0.280$. The redshifts of more distant objects ($z>2.34$) decrease; moreover, the rate
of decrease grows rapidly with recession (increasing $z$): $\dot{z}/H_0$ is
equal to $-0.98$, $-1.8$, $-8.4$, $-30$ at $z=4,\,5,\,10,\,20$ respectively.
For the ratio $\dot{z}/[H_0(1+z)]$, the growth is less pronounced. The
derivatives $\dot{z}$ and $\dot{z}/(1+z)$ are equal to zero at the same points,
and $\dot{z}/(1+z)$ reaches the maximum at $z=0.726$ that is smaller than the
maximum of $\dot{z}$.

Figure~\ref{fig:tcz} also presents a graph of the dimensionless derivative of the logarithm of the
apparent luminosity as a function of $z$. When $z=0$ this derivative equals $-2$, which reflects a
decrease in the solid angle of the source at the very beginning of the source's recession from the
observer. At small $z$ the rate of decrease grows slightly; at $z=0.726$ it has maximum negative
value, then decreases and becomes equal to zero at $z=13.2$. The apparent brightness of distant
objects should increase with redshift at $z>13.2$. The effect is stronger for more distant objects,
although it is unclear whether any radiating objects existed, since such redshifts correspond to
times $<326$ million years from the beginning of the expansion, less than a fraction $0.0238$ of the
current age of the universe.

It is most convenient to measure changes in all of these quantities when observing the
$L_\alpha$-forest, which corresponds to shifts of the $L_\alpha$ line in the spectra of distant
quasars due to gas clouds located along the path to them. These clouds can have peculiar radial
velocities relative to the Hubble flow that can affect observed shifts of the line. However, values
of these radial components most likely do not change significantly during the time between
observations if they are separated within some decades up to hundred of years. The luminosities of
objects do not change as well at least in average.

Despite the importance of this effect to test the theory, any possibility of observing it with
modern instruments would require a very long time interval between observations, from hundreds to
thousands years, since $\lambda(t_0)/\lambda(t)=1+z(t,t_0)$, then $\d\lambda_0/\lambda_0=\d
z/(1+z)$, and \be \d t_0=\frac {1}{H_0}\frac {\d z/(1+z)}{\dot {z}/[H_0(1+z)]}. \ee For example, if
we assume that the accuracy of measurements of a relative shift of lines is $\d\lambda/\lambda=\d
z/(1+z)=10^{-6}$ then for $z=4$, as seen from Fig.~\ref{fig:tcz}, $\dot{z}/[H_0(1+z)]\approx -0.2$
and $\d t_0\sim 14\cdot 10^9\cdot 10^{-6}/0.2=3\cdot 10^3$~yrs --- a time interval which is
insignificant on cosmological scales but longer than a human lifetime. For large $z$, the accuracy
of measuring the position of the lines is less, so that we would need major technological progress
to pursue such a method.

Liske \cite{Liske} estimates the possibility to detect the shift of lines due to cosmological expansion in
the spectra of various objects at different wavelengths for different cosmological models when
telescopes with ultra-large mirrors (40--60~m) become available, as is planned for the 2020s. It is
alleged that a 42-m telescope will be able to measure the shift with 4000 hour exposures separated
by 40 yrs. The article also provides an overview of previous work on the question of changing
redshift.

\section{The second inflation and the second horizon}

\subsection{Visible and invisible parts of the universe}

According to the theory of cosmological inflation, near the very beginning of the evolution of the
universe, space expanded exponentially. The standard theory, as follows from (\ref{eq:farfuture}),
predicts that a positive cosmological constant causes an acceleration of space starting from a
certain moment that leads to an exponential expansion, although at a much lower rate than during the
first inflation. This new expansion generates a new concept: a second horizon.

The equation of motion of a photon traveling to the observer (that is, to us) is $\chi=\eta_0-\eta$;
therefore, the place and time of its exit are related by the equality
$\chi_\e=\eta_0-\eta_\e<\eta_0$. So, the equation of motion can be rewritten as follows:
$\chi=\chi_\e+\eta_\e-\eta$. Therefore, the distance from the observer to the approaching photon is
\be \label{eq:lrsphot} l_\rs=l_\H^0a(\eta)(\chi_\e+\eta_\e-\eta). \ee The parameter $\eta$ is
limited. For $t=\infty$, it is equal to $\eta_\infty=4.4514$. The distance can only equal zero,
$l_\rs=0$, if $\chi_\e+\eta_\e<\eta_\infty$. Then there is another limitation on the ability to
observe objects in the universe: along with the first horizon there is a second one. The concept of
two horizons was introduced by V.\,Rindler \cite{Rindler} and discussed in a number of papers, for
example, in \cite{MMC}. Here their kinematic characteristics are derived within the Standard model.

The first horizon is called geometric (we recall that the physical horizon is the sphere of last
scattering at $z\approx 1000$), while the second horizon can be called the kinematic or dynamic
horizon. Other names are also used, borrowed from the terminology of the theory of black holes. The
geometric horizon is called the particle horizon, and the kinematic horizon is called the event
horizon. These names were introduced by Rindler.

At an arbitrary epoch, $\eta$, the first and second horizons are determined by the equations 
\be
\label{eq:psifirstsecond} \chi_\Hor=\eta=H_0\intl_0^a\frac {\d a}{a^2H}=\eta_*I_0(x,\beta),\nonumber\ee
\vspace{-0.5cm}\be\chi_\Horr=\eta_\infty-\eta=H_0\intl_a^\infty\frac {\d a}{a^2H}=\eta_*[I_0(\infty,\beta)-I_0(x,\beta)]. \ee 
\begin{figure}[tbp] \centering
\includegraphics[width=.45\textwidth]{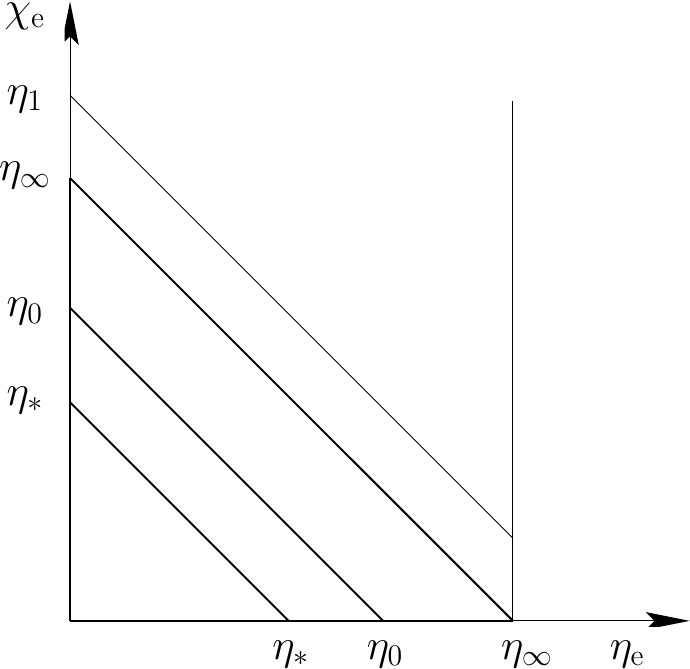} 
\caption{\label{fig:horizonn} Visible and invisible parts of the universe.} 
\end{figure} 
In Figure~\ref{fig:horizonn} positions of the geometric horizon are indicated on the ordinate axis. The
lines corresponding to this horizon are parallel to the abscissa. They rise with time, reflecting
expansion of the horizon. The second, kinematic horizon is shown by a straight line connecting the
abscissa and the ordinate, which are equal to $\eta_\infty$, while its specific position corresponds
to the time on the abscissa axis. The paths of photons coming toward us are represented by straight
lines parallel to this straight line. Photons can start their journey from any point on the
trajectory. Photons, for which $\eta_\e+\chi_\e<\eta_\infty$, that is, moving along straight lines
lying below the straight line specified above, sooner or later will reach a place where the observer
is located. For example, Figure~\ref{fig:horizonn} shows the paths of photons that have reached our
position at time $\eta_*<\eta_0$ and at the current epoch $\eta_0$. If
$\eta_\e+\chi_\e>\eta_\infty$, then photons with such coordinates never reach our location.
According to the equality $\eta_\e+\chi_\e=\eta_\infty$, it would seem that the photon still must
reach the observer at least over an infinite time, but even that is impossible.

From behind the first horizon, the radiation has not yet reached the observer. The second horizon
separates the region of times and locations from which radiation cannot reach the observer, since
the photons coming from there are moving away from the observer. This occurs because space expands
at speeds higher than the speed of light, and these speeds increase with time. At the current time,
we can see objects in the universe up to redshifts $z\approx 10$, but this corresponds to the past.
We will never see objects located now at redshifts $z\geq 1.725$.

Indeed, if a photon is now emitted toward us from a place with coordinate $\chi_\o$, then the
distance to it at moment $\eta$ will be $l_\ph=l_\H^0a(\eta)(\chi_\o+\eta_0-\eta)$. This distance
can become equal to zero at $\eta=\chi_\o+\eta_0$, and it must be the case that
$\chi_\o+\eta_0<\eta_\infty$. Thus, the boundary of the coordinate $\chi_\o$ for photons emitted now
is $\chi_\lim^0=\eta_\infty-\eta_0=1.13$. The values of $x_\lim^0=3.23$, $z_\lim^0=1.725$, and
$l_\lim^0=4.84$~Gpc$=l_\Horr^0$ correspond to this coordinate. The sphere of such a radius is the
current second horizon. Thus, the radiation from the points now located at distances of $4.84$~Gpc
from us will never reach us, even in the infinitely remote future. 
\begin{figure*}[tbp] \centering %
\includegraphics[width=0.99\textwidth]{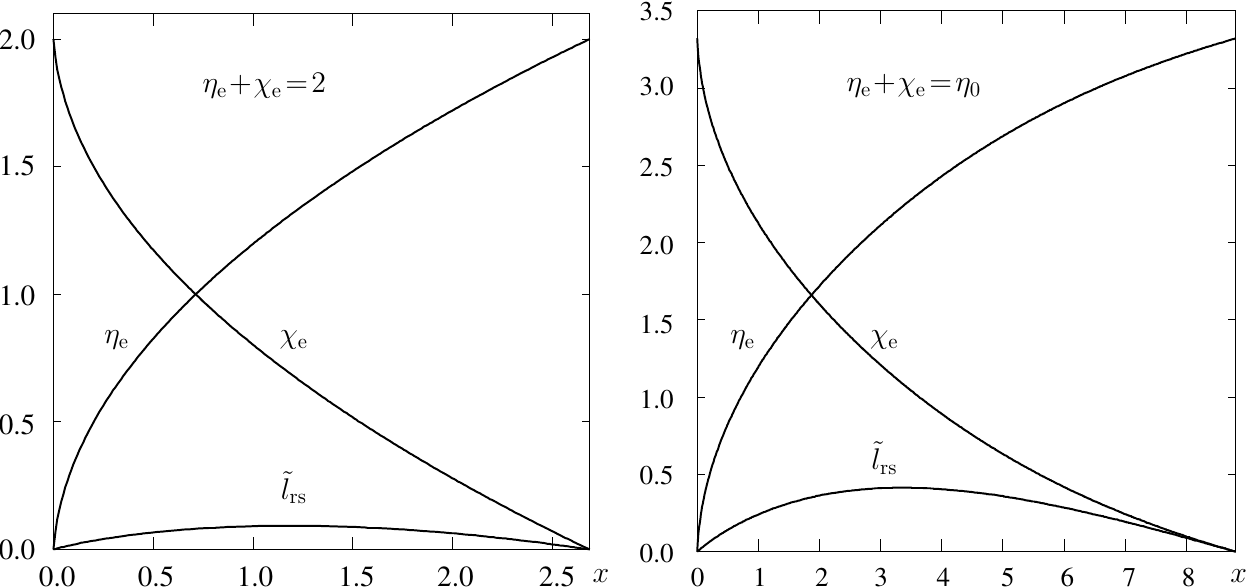} \caption {\label{fig:lphotet20} The paths of
photons with arrival time at epochs: $\eta=2$ ({\it left}) and $\eta_0=3.3224$ ({\it right}).} 
\end{figure*} 
\begin{figure*}[tbp] \centering 
\includegraphics[width=0.99\textwidth]{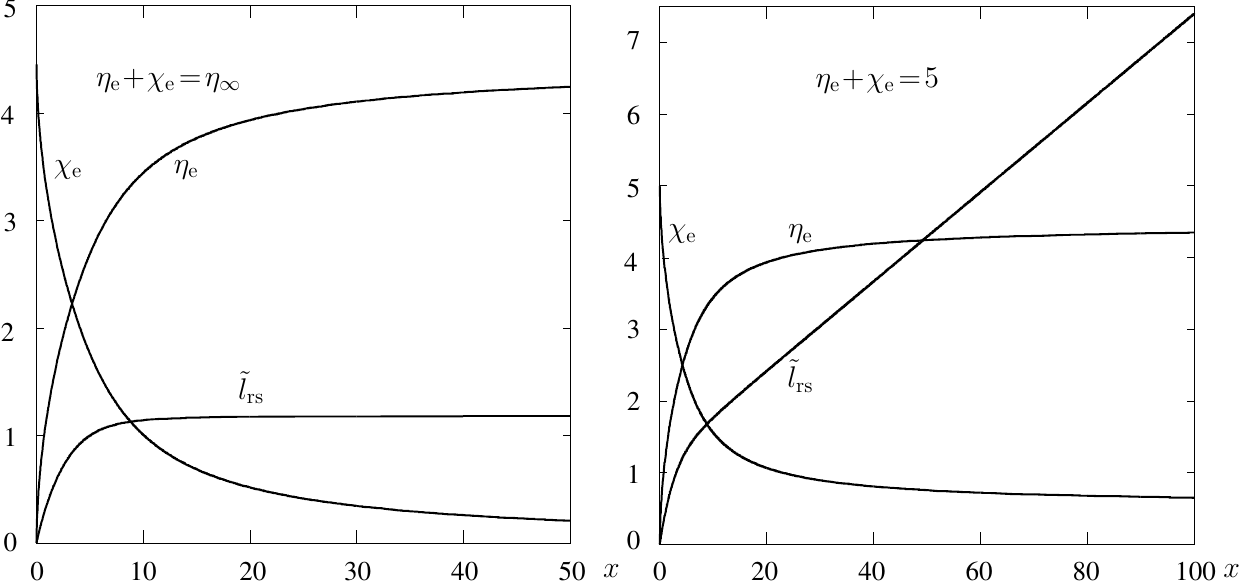} 
\caption {\label{fig:lphotetinf5} The paths
of photons with ``arrival'' time at epochs: $\eta_\infty=4.4514$ ({\it left}) and $\eta=5$ ({\it
right}).} \end{figure*} 
Figure~\ref{fig:lphotet20} shows distances $l_\rs$ to the photons arriving
at the observer at the epoch when $\eta=2$ (Figure~\ref{fig:lphotet20} {\it left}), and at the
current epoch (Figure~\ref{fig:lphotet20} {\it right}). In Figure~\ref{fig:lphotetinf5} these
distances are given as a function of values of $x$ for cases where the sum of the coordinates of
time and location of the photon emission is equal to $\eta_\infty$ (Figure~\ref{fig:lphotetinf5}
{\it left}) and larger than that (Figure~\ref{fig:lphotetinf5} {\it right}). These figures also show
curves reflecting the relationship between the time $\eta_\e$ and the location $\chi_\e$ of the
photon emission.

Generally speaking, if a photon is emitted at a point where the expansion speed is greater than the
speed of light, this does not necessarily mean that it will not reach us. Cosmological expansion
occurs in the same way with respect to all points of space, and it starts after a period of
inflation with a very high speed (formally infinite, according to formula (\ref{eq:Hofx}), which
defines the Big Bang), although in the beginning the expansion was slowing down. A photon emitted
from far away, where the speed of expansion is large but closer than the horizon, still comes to us,
because it gradually moves into layers of space expanding at a slower and slower rate. At some point
its velocity toward us becomes zero, and then becomes negative, that is, it begins to approach us.
However, it takes a long time for the photon to reach us. Consider, for example, a galaxy observed
by us now at redshift $z=3$: according to Table 2, it moves away from us with a speed of
$H_0l=c\tl=1.51 c$, and earlier its speed was greater. At the same time, its radiation traveled to
us for $11.5$ billion years, that is, we see this galaxy as it was in the distant past, when neither
the Earth, nor even the Sun, existed (but the galaxies and stars of the previous generations had
already formed).

Figure~\ref{fig:lphotet20} shows that the distance of a photon emitted sufficiently early at first
increases, which means expansion with a speed greater than the speed of light, faster than the
photon speed. From the point where the distance reaches a maximum, the photon begins to approach and
finally arrives at our location. However, as shown in Figure~\ref{fig:lphotetinf5}, this is not
possible if $\eta_\e+\chi_\e\geq\eta_\infty$, even if the equality is true.
Figure~\ref{fig:lphotetinf5} {\it left} shows that a photon emitted at the second horizon, and which
then travels along it, would not arrive at the observer after an infinite time; in fact, the photon
only recedes along with the horizon. After an infinite time, such a photon will be at a distance
$l_\Lambda\approx 5.0$~Gpc, since the factor $\eta_\infty-\eta$ in the formula (\ref{eq:lrsphot}) at
$\chi_\e+\eta_\e=\eta_\infty$ tends to zero if $t\to\infty$, while the factor $a(\eta)\to\infty$,
but their product remains finite. A photon emitted at $\eta_\e+\chi_\e<\eta_\infty$ may, after a
very long time, reach the current location of our civilization, but one emitted at
$\eta_\e+\chi_\e>\eta_\infty$ will only recede from us, eventually exponentially fast. The reason
for this is the accelerated expansion of space. Thus, galaxies located on the second horizon and
behind it will forever disappear from our field of view. These statements follow from the formulas
given below.

\subsection{Distances, velocities, and accelerations of horizons}

Distances to horizons at an arbitrary epoch $\eta$ according to equations (\ref{eq:psifirstsecond})
are defined by the formulas: \beq \label{eq:lHorxbetdef} & \strut\disp
l_\Hor=l_\H^0a(\eta)\eta=l_\Lambda xI_0(x,\beta), & \\ \label{eq:lHorrxbetdef} & \strut\disp
l_\Horr=l_\H^0a(\eta)(\eta_\infty-\eta)= l_\Lambda x[I_0(\infty,\beta)-I_0(x,\beta)]. & \eeq The sum
of the horizon conformal space coordinates is constant at all times, and the sum of the distances to
them is proportional to the scale factor. Both horizons expand. The speed of the geometric horizon
exceeds by the speed of light the velocity of the position where the horizon is located at time
$\dot{l}_\Hor=l_\H^0\dot{a}\eta+l_\H^0a\dot{\eta}=Hl_\Hor+c$. It expands at an accelerating rate. In
contrast, the velocity of the kinematic horizon is less than the speed of its location by the speed
of light: $\dot{l}_\Horr=l_\H^0\dot{a}(\eta_\infty-\eta)-l_\H^0a\dot{\eta}=Hl_\Horr-c$, and its
expansion slows down.

Asymptotes of distances to horizons and their velocities at $t\to\infty$, $a\to\infty$, $z\to -1$
are determined by taking into consideration that $I_0(\infty,\beta)=0.42880$ and $\disp
I_0(\infty,\beta)-I_0(x,\beta)\sim \frac{1}{x}\left(1-\frac {1}{8}\frac{\beta}{x^3}\right)$: 
\be\label{eq:asdistGhor}l_\Hor\sim l_\H^0\eta_*\,I_0(\infty,\beta)a= 5.9\cdot
10^{28}a\,\hbox {{\rm cm}}\sim 2.7\cdot 10^{28}e^{H_\Lambda t}\,\hbox {{\rm cm}}\to\infty\ee
\vspace{-0.5cm}\be l_\Horr\to\frac {c}{H_\Lambda}=1.56\cdot 10^{28}\,\,\hbox {{\rm cm}}=5.05\,\,\hbox
{{\rm Gpc}}. \ee\vspace{-0.5cm}

The current distance to the geometric horizon is $\disp
l_\Hor^0\!=\!l_\H^0\eta_0\!=\!3.32\,l_\H^0\!=\!4.39\cdot 10^{28}$~cm$\!= \!14.2$~Gpc. The velocity
near the horizon is $v_\Hor^0=c\eta_0=3.32c$, and the velocity of the expansion of the horizon is
$\dot{l}_\Hor^0=4.32c$. The horizon will cross $4.32$ light years in one year, which is equal to
$1.33$~pc, so that $1$~Gpc will be added to the current $14.2$~Gpc in $0.755\cdot 10^9$ years if the
speed of the horizon is equal to its current velocity, and in $0.741\cdot 10^9$ years if the
increase of the velocity is taken into account.

The current distance to the second horizon is
$l_\Horr^0\!=\!l_\H^0(\eta_\infty\!-\!\eta_0)\!=1.49\cdot 10^{28}$~cm$= \!4.84$~Gpc. The limit to
this distance coincides with the Hubble limit: $\disp
l_\Horr\to\frac{l_\H^0}{\sqrt{\Omega_\Lambda^0}}=\frac{c}{H_\Lambda}= 5.05$~Gpc. The current speed
of expansion of the location of this horizon is $c(\eta_\infty-\eta_0)=1.13c$, and the speed of
recession of the kinematic horizon from us is now $0.13c$.

The velocities of the horizons at an arbitrary moment and their asymptotics for $t\to\infty$ and
$x\sim 5.0\cdot e^{H_\Lambda t}\to\infty$ are \be \label{eq:vitHor} 
\dot{l}_\Hor\!=\!l_\H^0(\dot {a}\eta\!+\!a \dot {\eta})\!=\!Hl_\Hor\!+\!c\!=\nonumber\ee
\vspace{-0.5cm}\be\!c\left[\frac {\sqrt{1\!+\!\beta x\!+\!x^4}} {x}I_0(x,\beta)\!+\!1\right]\sim 
cxI_0(\infty,\beta),\ee
\vspace{-0.5cm}\be\label{eq:vitHorr} \dot {l}_\Horr=c\left[ \frac {\sqrt {1+\beta
x+x^4}}{x}[I_0(\infty,\beta)-I_0(x,\beta)]-1\right]\sim \frac {3}{8}\frac {\beta}{x^3}c. \ee 
It is interesting to note that all points with fixed coordinate $\chi$ begin (at the initial instant of
the expansion) to move away from each other at an infinite speed (according to (\ref{eq:Hofx}),
$\disp\dot{a}=\dot{x}/x_0=(H_\Lambda/x_0)\sqrt{1+\beta x+x^4}/x$). The geometric horizon begins to
expand with velocity $2c$, as does the Hubble distance, but the evolution of their velocities is
opposite (see formulas (\ref{eq:dotlH}) and (\ref{eq:asdistGhor})). For small $x$, the velocity
$\dot{l}_\Hor$ grows very fast, while the velocity $\dot{l}_\H$ rapidly decreases, so that at $x =
0.03$ they become equal to $2.50c$ and $1.56c$, respectively. The second horizon begins the
expansion, as do all ordinary points of space, with an infinite speed, which decreases very rapidly.

Accelerations have similar evolution: 
\begin{widetext}\beq \label{eq:accHor} & \strut\disp \ddot
{l}_\Hor\!=\!l_\H^0\left(\ddot {a}\eta\!+ \!2\dot {a}\dot {\eta}\!+\!a\ddot {\eta}\right)\!=\!\frac
{\ddot {a}}{a}l_\Hor \!+\!l_\H^0\left(2\dot {a}\frac {H_0}{a}\!-\!a\frac {H_0}{a^2}\dot {a}\right)\!
=\!-\frac {4\pi G}{3}\rho_\g l_\Hor\!+\!cH\!= & \nonumber \\ & \strut\disp =cH_\Lambda\left[\frac
{\sqrt {1+\beta x+x^4}}{x^2}+ \frac {x^4-\beta x/2-1}{x^3}I_0(x,\beta)\right]\sim cH_\Lambda
I_0(\infty,\beta) x, & \\ \label{eq:accHorr} & \strut\disp \ddot {l}_\Horr=l_\H^0\left[\ddot
{a}(\eta_\infty- \eta)-2\dot {a}\dot {\eta}-a\ddot {\eta}\right]=\frac {\ddot {a}}{a}l_\Horr-
l_\H^0\left(2\dot {a}\frac {H_0}{a}-a\frac {H_0}{a^2}\dot {a}\right)= & \nonumber \\ & \strut\disp
=-\frac {4\pi G}{3}\rho_\g l_\H^0 \frac {x}{x_0}\frac {I_0(\infty,\beta)-I_0(x,\beta)}
{(\Omega_\r^0\Omega_\Lambda^0)^{1/4}}-cH= & \nonumber \\ & \strut\disp =cH_\Lambda\left[\frac
{x^4-\beta x/2-1}{x^3} [I_0(\infty,\beta)\!-\!I_0(x,\beta)]\!-\!\frac {\sqrt {1+\beta x+x^4}}{x^2}
\right]\sim -\frac {9}{8}\frac {\beta}{x^3}cH_\Lambda. & \eeq\end{widetext} At $x=x_0$, we obtain the current
values of the velocities (see above) and accelerations: $\ddot{l}_\Hor^0=3.45H_\Lambda c=19.9\cdot
10^{-8}$\,\,cm/s$^2$, $\ddot{l}_\Horr^0=-1.95\,H_\Lambda c=-11.3\cdot 10^{-8}$\,\,cm/s$^2$.
\begin{figure*}[tbp] \centering 
\includegraphics[width=0.99\textwidth]{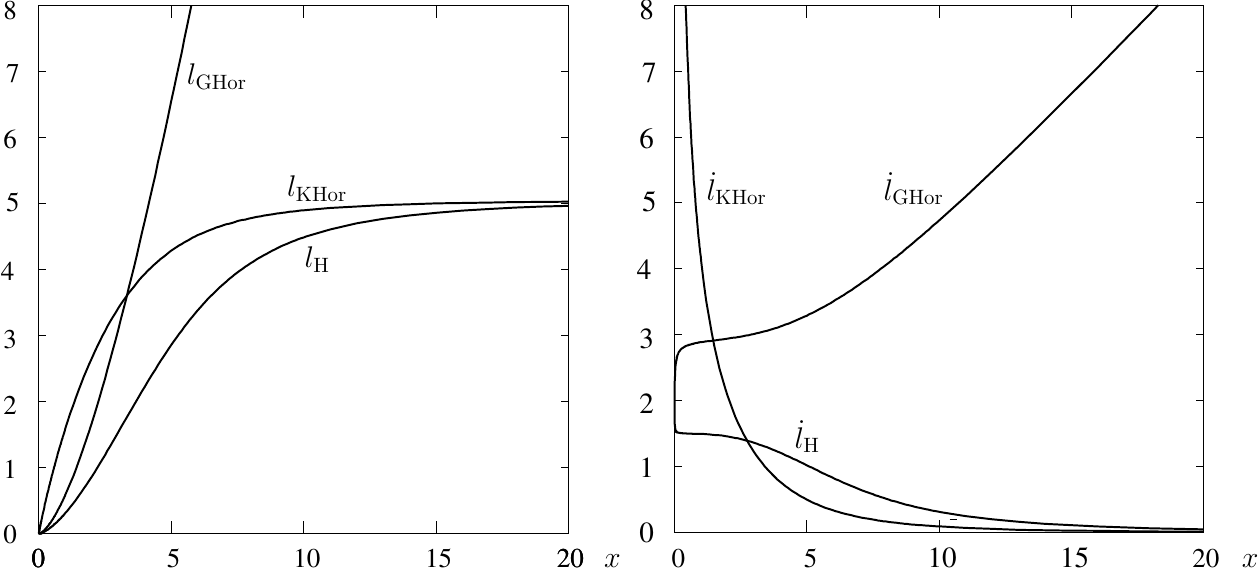} \caption {\label{fig:ldotlhgk} {\it Left:} Hubble
distance and distances to the horizons in Gpc. {\it Right:} Speeds of change of distances in units
of the speed of light. ({\it right}).} \end{figure*} 
\begin{figure*}[tbp] \centering %
\includegraphics[width=0.99\textwidth]{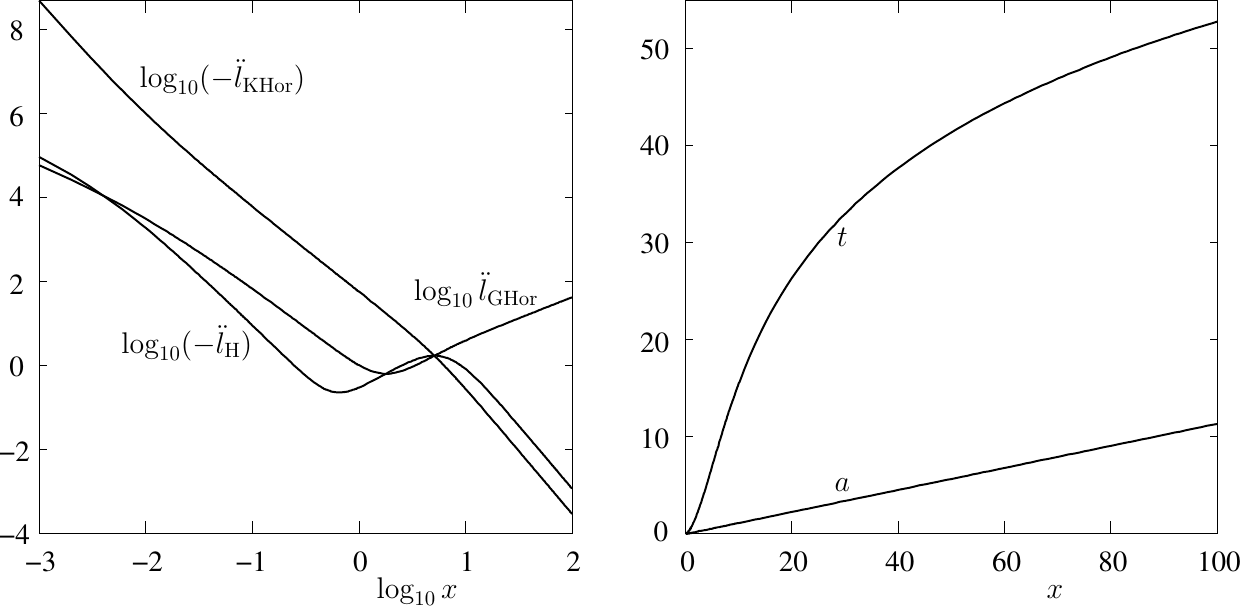} \caption {\label{fig:xdothgkat} {\it Left:}
Accelerations of the horizons and the Hubble distance. {\it Right:} Values of $a$ and $t$ as
functions of $x$.} 
\end{figure*}

The speed of the first horizon increases, while that of the second horizon decreases. During the
entire period of action of the cosmological acceleration ($6.5\cdot 10^9$ years), the velocity of
the first horizon increased from $3.31\,c$ (by the value of $\eta$ for $\rho_\g=0$ in Table 1) to
the current velocity of $4.32\,c$, and the speed of the second horizon decreased from $0.479\,c$ to
$0.129\,c$.

Figure \ref{fig:ldotlhgk}, {\it left} presents the distances to the horizons, and
Figure~\ref{fig:ldotlhgk}, {\it right} shows their velocities as a function of the parameter $x$.
The figures give the same values for the Hubble distance. All distances are given in Gpc, and
velocities are indicated in units of the speed of light. At first, until $\eta_\infty-\eta>\eta$,
the distance to the second horizon is greater than that to the first horizon. The horizons
intersected when $\eta_{\rm crs}=\eta_\infty/2=2.23$, $x_{\rm crs}=4.08$, $z_{\rm crs}=1.677$  at an
epoch $t_{\rm crs}=3.93$ billion years from the beginning, that is $t_0-t_{\rm crs}=9.80$ billion
years ago (earlier than the acceleration began), when the distance to the horizons was $3.58$~Gpc.
Prior to this, the first horizon determined the initial possibility to make observations (if there
were observers at that time). Since then, the second horizon has become closer. Note, however, that
the horizon effects differ. The first horizon (in fact, not it, but the physical horizon) limits the
spherical region of space in which one can observe the past history of the universe, while the
second defines those areas of information that will never reach the observer.

Figure \ref{fig:xdothgkat}, {\it left} plots the accelerations of the horizons and Hubble distance,
measured in units of $cH_\Lambda$, on a logarithmic scale. Only the acceleration of the second
horizon is a monotonic function; $\ddot{l}_\Hor$ has a minimum, while $\ddot{l}_\H$ has both a
minimum and a maximum. The figures show a linear increase of the acceleration $\ddot{l}_\Hor$ with
$x$, and the equality of the rates of decrease of the accelerations $\ddot{l}_\H$ and $\ddot{l}$
according to the asymptotes (\ref{eq:acclH}) and (\ref{eq:accHorr}). Figure~\ref{fig:xdothgkat},
{\it right} shows the relationship of the scale factor $a$ and cosmological time $t$ with the
parameter $x$.

\subsection{Connection with extraterrestrial civilizations}

Suppose that at the current epoch ($t=t_0,\,\,\eta=\eta_0$) humans emit a radio signal in some
direction. The distance to it increases; for a value of the time coordinate $\eta$ the distance will
be equal to $l_\ph=l_\H^0a(\eta)(\eta-\eta_0),\,\,\eta\geq\eta_0$. Its speed includes both the speed
of expansion and the speed of light: \be \label{eq:speedphot} \dot {l}_\ph=l_\H^0\dot
{a}(\eta-\eta_0)+l_\H^0a\dot {\eta}=Hl_\ph+c. \ee For brevity, we omit the factor $l_\H^0$, which
means that we use distances measured in units of the modern Hubble distance. On the way, the signal
passes by objects with fixed spatial coordinates $\chi_\O=\eta_0-\eta_\O$. Distances to these
objects grow only due to the cosmological expansion, that is, increasing scale factor:
$\tl_\O=a(\eta)(\eta_0-\eta_\O)$. The signal catches up with these objects when their distances from
us become equal, which occurs at the moment $\eta_\mt$, when $\eta_\mt-\eta_0=\eta_0-\eta_\O$,
$\eta_\mt=2\eta_0-\eta_\O$; therefore, $\tl_\ph=\tl_\O=a(2\eta_0-\eta_\O)(\eta_0-\eta_\O)$. Since
$\eta_\mt$ cannot exceed $\eta_\infty$, the signal can meet for a finite (although, perhaps, very
large) time only those objects whose coordinate satisfies $\chi_\O<\eta_\infty-\eta_0=1.13$. This
implies that the coordinate has the same boundary as a photon traveling toward us. This boundary is
the second horizon (see above), and $\eta_\O>\eta_\lim=2\eta_0-\eta_\infty=2.19$. 
\begin{table}[tbp]
\centering \caption{\label{tab:5} Objects along the signal path.} \smallskip\vspace{3mm}
\begin{tabular}{cccccc} 
\hline
$\eta_\O$ & $x_\O$ & $a_\O$ &$z_\O$ & $l_\O^0$ Gps & $t_\O$ \\ \hline
$1.99$ & $2.67$ & $0.303$ & $2.30$  & $5.69$ & $2.89$ \\
$2.19$ & $3.23$ & $0.367$ & $1.73$  & $4.84$ & $3.83$ \\ 
$2.39$ & $3.87$ & $0.440$ & $1.28$  &$3.98$ & $4.95$ \\ 
$2.59$ & $4.60$ & $0.523$ & $0.913$ & $3.12$ & $6.29$ \\ 
$2.79$ & $5.46$ &$0.620$ & $0.613$ & $2.27$ & $7.88$ \\
$2.99$ & $6.49$ & $0.737$ & $0.357$ & $1.41$ & $9.77$ \\
\hline
\end {tabular}\end{table}

\begin{figure*}[tbp] \centering 
\includegraphics[width=0.99\textwidth]{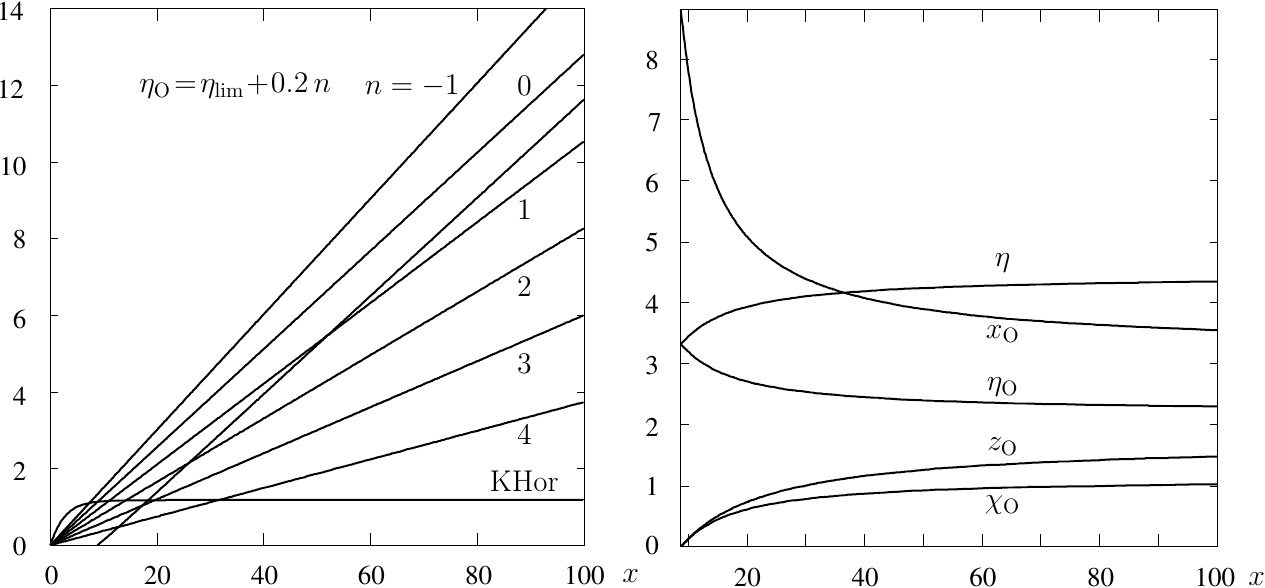} \caption {\label{fig:lphxooxox} {\it Left:} Path of
the signal to objects.~~~ {\it Right:} Objects reachable by the signal.} \end{figure*}

Figure~\ref{fig:lphxooxox}, {\it left} contains lines that plot the dependence of the distance
$l_\O$ on the coordinate $x$ up to six objects. Positions of the signal path and the kinematic
horizon are indicated as well. The objects are characterized by values of $\eta_\O=\eta_\lim+0.2n$,
$n=-1(1)4$. The corresponding values of $x_\O$, the scale factor $a_\O=a(\eta_\O)$, and the redshift
$z_\O$, as well as the current distances to these objects $l_\O^0$, are given in Table 5. It can be
seen from the figure that the emitted signal reaches the objects only when $n=1,2,3,4$. The signal
comes earlier to objects with larger values of $\eta_\O$, and hence smaller values of $z_\O$ and
$l_\O$. These objects are located closer to the position of the signal output at the moment of its
emission. The signal only catches up to objects with $n=3$ and $4$, for which the values of
$\eta_\O$ are equal to $2.79$ and $2.99$, respectively, before they cross the second horizon. The
condition of this is $\eta_0-\eta_\O<\eta_\infty-2\eta_\O+\eta_\O$, that is,
$\disp\eta_\O>\eta_\h=\frac{3\eta_0-\eta_\infty}{2}=2.758$. The distance between the object, which
is now almost on the second horizon (the current distance to it is $4.84$~Gpc and its redshift is
$1.72$), and the signal for $\tl_\O^0=\eta_\infty-\eta_0-\eps$, $\eta_\O=\eta_\lim+\eps$ is 
\be\label{eq:lomlph} l_\O\!-\!l_\ph\!=\!l_\H^0a(\eta)(\eta_0\!-\!\eta_\lim\!-\!\eps)\!-\!
l_\H^0a(\eta)(\eta\!-\!\eta_0)\!=\nonumber\ee
\vspace{-0.5cm}\be\!l_\H^0a(\eta)(\eta_\infty\!-\!\eps\!-\! \eta)\!\sim\!\frac
{c}{H_\Lambda}\frac {\eta_\infty\!-\!\eps\!-\!\eta} {\eta_\infty-\eta}, \ee since, according to eqs.
(\ref{eq:H0tI1xbeta}) and (\ref{eq:I0aprlrgx}), $\eta\sim\eta_\infty-\eta_*/x$, $\disp
a(\eta)=x/x_0\sim 1/ \left[\sqrt{\Omega_\Lambda^0}(\eta_\infty-\eta)\right]$ for $x\to\infty$,
$\eta_\infty-\eta\ll 1$. The difference (\ref{eq:lomlph}) tends to zero for
$\eta\to\eta_\infty-\eps$ if $\eps>0$. Thus, in agreement with Figure~\ref{fig:lphxooxox}, {\it
left}, the signal will still reach a given object if the object is located at least slightly closer
than the second horizon. The time that the signal needs to meet the object is $t\sim\ln(1/\eps)$. If
the object is located on the horizon ($\eps=0$), there remains an insurmountable distance
$c/H_\Lambda=5.05$~Gpc. The distance between the signal emitted now and objects currently located
behind the second horizon will only increase with time. In addition, it will increase asymptotically
as an exponential function. These objects are carried away by the exponential expansion, that is, by
the repulsion of the dark energy. In models without repulsion the second horizon does not appear.

Figure \ref{fig:lphxooxox}, {\it right} shows the dependences of the coordinates $x_\O$, $\eta_\O$
and $\chi_\O=\eta_0-\eta_\O$, as well as redshifts $z_\O$  of objects that the signal will reach at
the time corresponding to its coordinate $x$. The figure plots also the dependence of the time
coordinate $\eta$ on $x$. The signal emitted now will reach the second horizon when 
\be\label{eq:pathtoHor} l_\ph=l_\H^0a(\eta_\h)(\eta_\h-\eta_0)=l_\Horr=l_\H^0a(\eta_\h)
(\eta_\infty-\eta_\h),\nonumber\ee
\vspace{-0.5cm}\be\eta_\h=\frac {\eta_0+\eta_\infty}{2}=3.89. \ee The corresponding values
are $x_\h=18.3$, $l_\ph=l_\Horr=5.02$~Gpc, and $t_\h=24.9$~Gyr. At this time an object with initial
coordinate $x_\O=5.30$ will approach the horizon. This coordinate corresponds to $a_\O=0.601$,
$z_\O=0.663$, and an initial distance to us of $l_\O^0=l_\H^0(\eta_0-\eta_\O)=1.45$~Gpc, because if
$l_\O=l_\ph=l_\Horr$ then $\eta_0-\eta_\O=\eta_\h-\eta_0=\eta_\infty-\eta_\h$ and
$\eta_\O=(3\eta_0-\eta_\infty)/2=2.76$.

The signal sent by us can reach distances only up to $\approx$5~Gpc to have any hope to get a reply.
Exponential expansion of space entrains radiation both going away from us and directed toward us.
Nevertheless, $5$~Gpc is a very large distance, and inside the sphere of such a radius there are
many galaxies. If the signal hits a planet populated by intelligent creatures who have reached an
advanced stage of civilization, they can receive it, understand, determine the direction from what
it came, and reply. Then their signal will approach that place where humans were when we sent the
first signal. The distance of their signal to us will change according to the formula
$\tl_\ret=a(\eta)(\eta_\mt-\eta_0-\eta)=a(\eta)(\eta_0-\eta_\O-\eta)=
a(\eta)(3\eta_0-2\eta_\O-\eta)$. Any reply will arrive at Earth at time
$\eta_\ret=3\eta_0-2\eta_\O$. If we want a reply to arrive at time $\eta_\ret^0$,
$\eta_0<\eta_\ret^0<\eta_\infty$, then such a civilization should have
$\eta_\O>(3\eta_0-\eta_\ret^0)/2$ and $\chi_\O<(\eta_\ret^0-\eta_0)/2$. In an extreme case, if we
assume that $\eta_\ret^0=\eta_\infty$, then the condition $\eta_\ret<\eta_\infty$ imposes a
restriction on the coordinate $\eta_\O$: $\disp\eta_\O>(3\eta_0-\eta_\infty)/2$. This restriction
coincides with the condition that the signal reaches the object (civilization) before the latter
reaches the second horizon. The restriction on the spatial coordinate is as it was previously:
$\chi_O=\eta_0-\eta_O<(\eta_\infty-\eta_0)/2$.

It is clear that it makes sense to send a signal to objects located closer than several dozen light
years, otherwise any possible reply would take too long. Undoubtedly, it will be necessary to limit
the search within our Galaxy and even the immediate vicinity of the solar system. Even in this case
the signals must either be sent in a very narrow cone, or they should be sufficiently energetic so
that they can be received at a greater distance.

Although the above arguments have a purely theoretical or even academic character, they establish
restrictions on the limits imposed by the model. They can be related either to epochs when our
civilization on the Earth has not existed yet, or has not been able to realize connections with
other civilizations, or to the epochs when the Sun and Earth will no longer exist in their current
form. However, these same arguments apply to any arbitrary location in the universe and to
civilizations that may arise and prosper at any time.

\section{Conclusion}

In this paper we have summarized results of the Standard model that reveal some of its quantitative
properties. After a brief excursion into the history of creation of cosmological models, we have
presented the two Friedmann-Lema\^{i}tre equations, which describe a uniform and isotropic universe,
and have restated the definitions of the critical values and five cosmological distances. Based on
the compatibility condition of two equations and on the equations of state, we have derived the laws
of the change with time of the mass density of four noninteracting components: the density of matter
decreases as the third order of the scale factor and the density of radiation and neutrinos as the
fourth order, while the dark energy density is unchanged. These laws provide solutions of the
cosmological equations in quadratures.

The equations are specified for the flat space-time model. Using the parameters of this model
obtained from observations: Hubble constant, $H_0=70$~km/s/Mpc, temperature of the cosmic background
radiation, $T_0=2.7727$~K, dark energy fraction of the total cosmological mass density,
$\Omega_\Lambda^0=0.72$, we have determined the current Hubble distance $l_\H^0=4.28$~Gpc, the
critical density $\rho_\cc=9.2\cdot 10^{-30}$~g/cm$^3$, and the fractional contributions to
$\rho_\cc$ of the radiation, $5\cdot 10^{-5}$, six types of neutrinos $6.9\cdot 10^{-5}$, and
dust-like matter, $\approx 0.28$, which includes dark matter. They define the relationships of the
scale factor $a=1/(1+z)$ (and, thus, of the redshift $z$), the conformal dimensionless time
coordinate $\eta$, and the dimensionless parameter $x$ with cosmological time $t$. For the early and
late stages of expansion, simple and sufficiently accurate approximations of these relationships
have been obtained. We have shown that, in contrast to redshift, coordinates $\eta$ and $x$ are not
tied to a specific epoch of evolution. Several dimensionless parameters of the model have been
introduced that are also free from such binding, and their quantitative values have been derived.

It has been shown that major events in the evolution of the universe occurred near the beginning,
when the dominant carrier of the mass density transferred from radiation and neutrinos to ``dust''
--- corresponding to $z$ changing from $5500$ to $1000$ --- and then later when dark energy became
dominant, at $z$ = $0.7$ to $0.4$.

Dependencies of different types of distances have been calculated as functions of parameters $x$ and
$z$, as well as speeds of their changes. A difference between the cosmological redshift and
classical Doppler effect was stressed, which was explained by the fact that a shift of the frequency
of a photon occurs not only at the time of its emitting by a cosmical object, but at every point of
its path to the observer.

We have discussed the concepts of two horizons: geometric, inherent in any expanding model, and
kinematic, typical for models expanding with an acceleration. Distances to these horizons, along
with speeds and accelerations of evolution of these distances, have been derived as functions of
time. Current distances to the horizons are $14.2$~Gpc for the first horizon and $4.84$~Gpc to the
second horizon. The horizons crossed each other $9.8$ Gyr ago, when the distance to them was
$3.58$~Gpc.

The current acceleration of expanding space represents the most surprising value: at the Hubble
distance, where the expansion rate is equal to the speed of light, the acceleration is about $4$
$\Angs/$s$^2$. Such a value of the acceleration has been reached over the past $6.5$ billion years,
while it was zero at the beginning of the expansion. A limit on the acceleration for $t\to\infty$ at
the limit of the Hubble distance is $5.7$ $\Angs/$s$^2$. Even at the current horizon the
acceleration is only slightly higher, $\approx 20$ $\Angs/$s$^2$. However, tens of billions of years
into the future, the acceleration of the horizon and scale factor (indeed, all scales) will increase
over time exponentially with an exponent of $t/t_\Lambda$, where $t_\Lambda=16.5$~Gyrs. This means
that a second inflation will occur.

We have estimated the rate of change of redshifts and apparent luminosities of objects with
increasing age of the universe. For distant objects with $z> 13.2$, the apparent luminosity can grow
with time. However, detection of these effects requires very long time intervals between
observations, as well as significant improvements in the capabilities of observational instruments
in the future.

We have estimated distances across which our signal emitted from the Earth can reach
extraterrestrial civilizations and from which they can respond to us. These distances are quite
large, on the order of $5$~Gpc, so that they do not limit the possibilities of contact with other
civilizations in the universe.

The above discussion provides a quantitative description of various geometric and kinematic
properties of the Standard model that is currently considered to be an accurate description of the
universe. Future observationally driven revisions of the cosmological parameters such as the Hubble
constant would require updates to the exact values of the parameters that we have derived.
Nevertheless, as long as the cosmological constant dominates the current and future energy of the
universe, seemingly odd features, such as the presence of two horizons, will remain intact.

%

\end{document}